\documentclass[twocolumn]{aa} 

\usepackage{gensymb} 
\usepackage{amsmath} 
\usepackage{graphicx, xspace}
\usepackage{xcolor}
\usepackage{natbib}
\usepackage[varg]{txfonts}
\usepackage{subcaption}
\usepackage{comment}
\usepackage{ulem}
\usepackage{booktabs,caption}
\usepackage[flushleft]{threeparttable}
\usepackage{multirow}
\bibpunct{(}{)}{;}{a}{}{,} 

\usepackage[breaklinks, colorlinks, citecolor=blue, linkcolor=blue]{hyperref}
\makeatletter
\def\instrefs#1{{\def\scsep{\def\scsep{,}}\@for\w:=#1\do{\scsep\ref{inst:\w}}}}
\renewcommand{\inst}[1]{\unskip$^{\instrefs{#1}}$}
\newcommand{\texttable}[1]{{\textbf{#1}}}
\def\Bimf{$B_{\rm IMF}$}
\def\Pdsw{$P_{\rm d}$}
\def\Pimf{$P_{\rm IMF}$}
\def\Rin{$R_{\rm in}$}
\def\Rout{$R_{\rm out}$}

\def\ergs{\mbox{\,erg~s$^{-1}$}}

\newcommand{\Rp}{\ensuremath{R_{\text{p}}}}
\newcommand{\Bp}{\ensuremath{B_{\text{p}}}}
\newcommand{\Rmp}{\ensuremath{R_{\text{mp}}}}

\newcommand{\nsw}{\ensuremath{n_{\text{sw}}}}

\newcommand{\Mdotstar}{\ensuremath{\dot{M}_\star}}
\newcommand{\Mdotsun}{\ensuremath{\dot{M}_\sun}}

\graphicspath{{./}{./pics/}}

\begin{document} 

\title{MHD simulations of the space weather in Proxima b: \\ Habitability conditions and radio emission}

\titlerunning{}
\authorrunning{Main author et al.}

 \author{L. Pe\~na-Mo\~nino{\inst{iaa}} \and
         M. P\'erez-Torres{\inst{iaa,capa,euc}}  \and
         J. Varela\inst{uc3m,IFS}  \and 
         P. Zarka{\inst{lesia,usn}}
        }

 \institute{
 CSIC, Instituto de Astrof\'isica de Andaluc\'ia, 
Glorieta de la Astronom\'ia S/N, E-18008, Granada, Spain \label{inst:iaa} 
    \\
    \email{\href{mailto:lpm@iaa.es, torres@iaa.e}{lpm@iaa.es,torres@iaa.es}} 
   \and
   Center for Astroparticles and High Energy Physics (CAPA), Universidad de Zaragoza, E-50009 Zaragoza, Spain  \label{inst:capa} 
   \and
   School of Sciences, European University Cyprus, Diogenes street, Engomi, 1516 Nicosia, Cyprus
   \label{inst:euc}
   \and
 Universidad Carlos III de Madrid, E-28911, Legan\'es, Spain \label{inst:uc3m} 
  \email{\href{mailto:jvrodrig@fis.uc3m.es}{jvrodrig@fis.uc3m.es}}
     \and
\label{inst:IFS} Institute for Fusion Studies, University of Texas, Austin, TX 78712, United States
    \and
   LESIA, Observatoire de Paris, CNRS, PSL, SU/UPD, Meudon, France  \label{inst:lesia} 
    \and
    USN, Observatoire de Paris, CNRS, PSL, UO, Nançay, France \label{inst:usn} 
}

\date{version of \today}

\abstract
{The habitability of exoplanets hosted by M-dwarf stars dramatically depends
on the space weather, where the magnetic and ram pressure of the stellar wind, and the exoplanet magnetic field are the three main players. Those three parameters also likely drive the radio emission arising close to the planet. 
}
{
We aim at characterizing the magneto-plasma environment and thus the habitability of the Earth-like planet Proxima b, which is inside the habitable zone of its host M-dwarf star Proxima, when it is subject to average, calm space weather conditions, as well as to more extreme space weather conditions, e.g., a CME-like event.  We study the role of the stellar wind and planetary magnetic field, and their mutual orientation. We also
determine the radio emission arising  from the interaction between the
stellar wind of Proxima and the magnetosphere of its planet Proxima b, which
is relevant to guide radio observations aimed at unveiling planets.
}
{
We use the PLUTO code to run a set of 3D magneto-hydrodynamic simulations
focused on the space weather around planet Proxima b. 
We consider  both calm and space weather conditions for Proxima b, under three different scenarios:
(a) Proxima b subject to calm space weather in a  sub-Alfvénic regime, where the stellar wind
magnetic pressure dominates over the wind's ram pressure; (b) Proxima b is subject to calm space weather in a super-Alfvénic regime, where the ram pressure of the wind dominates, and a bow
shock is formed; and (c)  Proxima b is subject to a Coronal Mass Ejection event, when the dynamical and magnetic pressure of the stellar wind from its host star are increased enormously for a short period of time. 

}
{
 We find that if Proxima b has a magnetic field similar to that of the Earth ($B_{\rm p} = B_\oplus \approx 0.32$ G) or larger, the magnetopause standoff distance is large enough to shield the surface from the stellar wind for essentially any planetary tilt but the most extreme values (close to
  $90^{\degree}$), under a calm space weather. 
  Even if Proxima b is subject to more extreme space weather conditions, e.g., a CME event from its host star, the planet is well shielded by an Earth-like magnetosphere (\Bp $\approx B_\oplus$; $i \approx 23.5\degree$), or if it has tilt smaller than that of the Earth. Otherwise, the planetary magnetic field must be larger to shield the planet from particle precipitation on the surface.
  For calm space weather conditions, the radio emission caused by the day-side reconnection regions can be as high
  as 7$\times10^{19}$ \ergs\ in the super-Alfvénic regime, and is on average
  almost an order of magnitude larger than the radio emission in the sub-Alfvénic
  cases, due to the much larger contribution of the bow shock, which is not formed in the sub-Alfvénic regime.
    We also find that the energy dissipation at
  the bow shock is essentially independent
  of the angle between the planet's magnetic dipole and the incident stellar wind flow.
If Proxima b is subject to extreme space weather conditions, the radio emission is more than two orders of magnitude larger than under calm space weather conditions. This result yields expectations for a direct detection--from Earth--in radio of giant planets in close-in orbits, 
as they are expected to have magnetic fields large enough, so that their electron-cyclotron frequency exceeds the ionosphere cutoff.
  
}
{}


\maketitle

\nolinenumbers
\section{Introduction}
\label{sec:intro}

The space weather of exoplanets depends on the properties of the stellar wind,
mainly through its density ($n_{\rm sw}$), velocity ($v_{\rm sw}$), and
magnetic field (\Bimf), as well as on the magnetic field of the planet (\Bp).
If the exoplanet orbit is located in the inner part of the habitable zone,
extreme conditions, or the absence of a significant magnetic field can lead to
direct deposition of the stellar wind towards the exoplanet surface, thus
threatening the planet habitability  \citep{Varela2022SpWea}.  In the case of
the Earth, whose host star is a G-type star, its magnetic field is strong
enough to avoid the direct precipitation of the damaging particles and
high-energy radiation from the solar wind on the surface, even during the
largest coronal mass ejections (CMEs) observed \citep{Kilpua2019,Hapgood2019}.
CMEs are stellar eruptions produced by magnetic reconnections in the
stellar corona \citep{Low2001}, which expel a magnetic cloud of charged paticles moving at a few times the wind speed (thousands of km/s) \citep{Neugebauer}.
Extreme space weather events are not
exclusive of Sun-like stars, and have also been observed in M, K and F type
stars \citep{Khodachenko2007,Lammer2007}. 

The space weather around exoplanets cannot be directly compared to the case of
the Earth if the host star has characteristics different from the Sun
(stellar type, age, metallicity, magnetic field, rotation, etc.). If the dynamic pressure, \Pdsw, and
the magnetic pressure, \Pimf, of the stellar wind are large,  favourable
exoplanet habitability conditions necessarily  require an intrinsic exoplanet
magnetic field strong enough to prevent the direct precipitation of the
stellar wind on the exoplanet surface (e.g., \citealt{Airapetian2020} and
references therein).  Otherwise, if the exoplanetary magnetic field is not
strong enough, the exoplanet habitability will be hampered by the effect of
the stellar wind, as well as the depletion of the atmosphere, especially
volatile components such as water molecules \citep{Jakosky2015}, 
although the role of the planetary magnetic field versus atmospheric escape may be more complex than a simple shield \citep{Gronoff2020}.

Space weather conditions inside the stellar habitable zone depend on the
characteristics of the host star
(e.g., \citealt{Airapetian2020}).
For M dwarf stars ($< 0.5 M_{Sun}$) the habitable zone is between $0.03$ and $0.25$ au
\citep{Shields2016}.
The habitability conditions on exoplanets inside the habitable zone of M dwarf
stars are still an open issue. First, those exoplanets are likely to be
tidally locked \citep{Griesmmeier2004AA} and exposed to a strong radiation
from the host star \citep{Griessmeier2005AsBio,Scalo2007} as well as
persistent CME events. On the other hand, recent studies indicate that tidal
locking may constrain but not preclude the habitability conditions of
exoplanets \citep{Barnes2017}.
Second, space weather conditions also
change with the rotation rate of the star, because the magnetic activity and
the properties of the stellar wind generated by the star change \citep{Suzuki2013}.

The interaction of the stellar wind with an exoplanetary magnetosphere results in the formation of magnetospheric reconnection regions.  The dissipation of energy at the magnetopause or, in a super-Alfvénic regime, at the bow shock, leads to indirect, or direct electron acceleration (e.g, direct acceleration 
driven by a Dungey-like cycle;  \citealt{Dungey1961}). 
Those accelerated electrons precipitate towards the central magnetized body (the planet in the case of magnetospheric emission; most likely the star in the case of sub-Alfvénic star-planet interaction). Eventually, the precipitating electrons are responsible for the emission at cyclotron radio frequencies near the magnetized body. This radio emission happens either indirectly, via the loss-cone emission \citep{Wu1979}, where the emitting electrons are the ones reflected at their mirror points; or directly, 
e.g., via beam, or shell emission  (e.g. \citealt{Hess2008}).

The  mechanism behind this
radio emission is the electron-cyclotron maser instability \citep{Wu1979}, which is responsible for the
low-frequency emission seen from the interaction of the solar wind with
planets of the solar system with intrinsic magnetic fields, e.g., Jupiter
\citep{Kaiser,Zarka5}, where a fraction of the electrons energy is transformed
into cyclotron radio emission escaping from the magnetosphere.
Likewise, the radio emission detected from an exoplanet magnetosphere could
provide information of the exoplanet intrinsic magnetic field \citep{Hess}.
Unfortunately, the detection capability of present radio telescopes barely
distinguish the radio emission from exoplanets. Recently, different
observations have shown tentative detections of radio emission from confirmed
star-planet systems, e.g., Proxima - Proxima b , using the ATCA
\citep{PerezTorres2021}; Tau Boo - Tau Boo b, using the LOFAR telescope
\citep{Turner2021}; YZ Cet - YZ Cet b, using the VLA \citep{Pineda2023} and
the GMRT \citep{Trigilio2023}.  In addition, \citet{Vedantham2020} interpreted
LOFAR observations of the red dwarf GJ $1151$ as being due to the magnetic
star-planet interaction with a putative, yet undiscovered, exoplanet.
Unfortunately, none of the above cases have shown conclusive evidence of the
radio emission arising from star-planet interaction (Proxima, GJ~1151, YZ Cet)
or directly from the planet's magnetosphere (Tau Boo).

We carried out our study using the single fluid MHD code PLUTO in spherical 3D coordinates \citep{Mignone2007}. 
We have already applied this numerical framework to model global structures of the Hermean magnetosphere  
\citep{Varela2015,Varela2016v120,Varela2016v122,Varela2016v129},
the effect of an interplanetary CME-like space weather conditions on the Earth magnetosphere \citep{Varela2022AA, Varela2023}, slow modes in the Hermean magnetosphere \citep{Varela2016v125, Varela2022AA} as well as the radio emission from the Hermean and exoplanetary magnetospheres \citep{Varela2016,Varela2018,Varela2022SpWea,Mishra2023}.

Proxima Centauri is an M-dwarf star, and the closest to our Sun. It is
fully convective and slowly rotating ($P_{\rm rot} \sim 84$ d;
\citealt{Mascareno2016}), with a large-scale magnetic field estimated between
$\sim 200$ G \citep{Klein2021} and up to $\sim 650$ G \citep{Reiners2008},
which seems to have a magnetic cycle of about 7 yr \citep{Mascareno2016}.
Since Proxima b is in the habitable zone of its host star, 
at a separation of 0.049 au and with an excentricity of 0.02,
the stellar activity level is expected to have a
crucial influence on the actual planet habitability  (e.g.,
\citealt{Cohen2014}).  Although Proxima is a slow rotator and, therefore, is
less active than other, faster rotating, M dwarf stars, it has still
significant levels of activity.  M dwarfs exhibit in general high levels of
activity, with fast-rotating M dwarfs being more active than slow-rotating
stars, such as Proxima.  For example, \citet{Ribas2016} found that Proxima b
receives 30 times more extreme-UV radiation than Earth and 250 times more
X-rays. We note, though, that this is essentially due to the large difference
in  the star-planet separation of Proxima-Proxima b compared to that of the
Sun-Earth, which is over 20 times larger.  If corrected for the different
distance dilution factors, the extreme UV and X-ray irradiation of Proxima b
would be factors of $\sim$14 and $\sim$2 smaller, respectively, compared to
that of the Earth, as the quiescent x-ray and extreme-UV luminosities of
Proxima are smaller than those of the Sun.

The planet Proxima b is an Earth-like planet that lies inside the habitable zone of
its host M-dwarf star Proxima Centauri \citep{Anglada-Escude2016}. It may be
subject to the effects of CMEs and stellar energetic events, as indicated by
the detection of a type IV burst \citep{Zic2020}, whose occurrence is strongly
associated with such energetic phenomena. MHD simulations also indicate the importance of space weather conditions on the habitability of
Proxima b, which overall faces a more adverse environment than the Earth
\citep{Garraffo2016,GarciaSage2017,Garraffo2022}. In fact, the 
stellar wind dynamic pressure at the orbit of Proxima b can be up to three
orders of magnitude higher compared to that of the Earth, particularly during
extreme space weather conditions. If calm space weather conditions at the
Earth and Proxima b are compared, the wind
density  is around 10 times larger and its velocity twice as large at
Proxima b. Likewise, the IMF intensity is more than 10 times bigger.  On the
other hand, space weather conditions may affect the habitability of Proxima b,
as far as the shielding provided by its planetary magnetosphere is concerned,
so as to avoid the sterilizing effect of the stellar wind on the surface (see
\citealt{Varela2022SpWea} and references therein).  Finally, 3D Global Climate
Modelling of the atmosphere and water cycle of Proxima b show that its
habitability is possible for a very broad range of atmospheric pressures and
compositions, and the presence of surface liquid water requires either a large
surface inventory of water (a global ocean able to resupply H2O to the dayside
by deep circulation) or an atmosphere with a strong enough greenhouse effect
that increases surface temperatures above the freezing point of water
\citep{Turbet2016}.

Finally, we note that Proxima b is also a primary target for radio studies,
and observations with the Australia Telescope Compact Array have suggested that
the observed coherent bursting  radio emission may arise from (sub-Alfvénic)
magnetic star-planet interaction \citep{PerezTorres2021}, although later
magnetohydrodynamic simulations suggest that Proxima b is likely to lie in
the super-Alfvénic regime \citep{Kavanagh2021}.  

In this paper, we present a study of both calm and extreme space weather conditions applied to
the closest system hosting an Earth-like exoplanet, the Proxima - Proxima b
system. In particular, we study the effect of the planetary magnetic field,
and of the angle it makes with the incident stellar wind flow, on the
habitability of Proxima b. 

We also study the radio emission that arises from the electrons accelerated in
the magnetic reconnection regions. 

Our present study follows a methodology similar to that carried out in previous studies dedicated to analyse the effects of space weather conditions on exoplanet habitability and radio-emission generation from planetary magnetospheres \citep{Varela2022AA,Varela2022SpWea}.  
We describe the numerical code PLUTO in Sect.~\ref{sec:pluto}. 
In Sect.~\ref{sec:simulations}, we apply the PLUTO code to study and discuss the 
magneto-plasma environment of Proxima b.
We provide a detailed comparison between the MHD simulations performed in this work against those published in previous ones on Proxima Cen. We also highlight the qualitative differences of our approach compared to those in other works in Sect. \ref{sec:simulations}. We also emphasise the quantitative differences of our work (Table \ref{tab:pluto-params}) versus the works by others (see Table~\ref{tab:params-comparison}). 
In Sect.~\ref{sec:magnetopause}, we discuss the habitability conditions of Proxima b, using the magnetopause standoff distance as a proxy, and in Sect.~\ref{sec:radio}, we determine the radio emission that is expected to arise close to Proxima b.

Finally, in Sect.~\ref{sec:summary} we summarise the main results and conclusions from the paper.

\section{Magnetohydrodynamic numerical simulations with PLUTO}
\label{sec:pluto}

PLUTO is an open-source, full 3D MHD code
in spherical coordinates that computes the evolution of a single-fluid,
polytropic plasma in the nonresistive and inviscid limit \citep{Mignone2007}. 
The equations solved in our model are the mass, momentum, magnetic field, and
energy conservation equations, for an ideal gas.  We show below the equations
in conservative form:

\begin{equation}
\label{Density}
\frac{\partial \rho}{\partial t} + \mathbf{\nabla} \cdot \left( \rho\mathbf{v} \right) = 0
\end{equation}

\begin{equation} 
\label{Momentum}
\frac{\partial \mathbf{m}}{\partial t} + \mathbf{\nabla} \cdot \left[ \mathbf{m}\mathbf{v} - \frac{\mathbf{B}\mathbf{B}}{4\pi} + I \left(P + \frac{\mathbf{B}^2}{8\pi} \right)  \right]^{T} = 0 
\end{equation} 

\begin{equation}
\label{B field}
\frac{\partial \mathbf{B}}{\partial t} + \mathbf{\nabla} \times \mathbf{E} = 0
\end{equation} 

\begin{equation}
\label{Energy}
\frac{\partial E_{t}}{\partial t} + \mathbf{\nabla} \cdot \left[ \left( \frac{\rho \mathbf{v}^{2}}{2} + \rho e + P \right) \mathbf{v} + \frac{\mathbf{E} \times \mathbf{B}}{4\,\pi}  \right] = 0, 
\end{equation}

where $\rho$ and $\mathbf{v}$ are the plasma density and velocity, $\mathbf{m}
= \rho\,\mathbf{v}$ is the momentum density, $P$ is the thermal pressure of
the plasma, $\mathbf{B}$ is the magnetic field,  $E_t = \rho \mathbf{v}^2/2 +
\rho e + B^2/8\pi$ is the total energy density, $\mathbf{E} = -
(\mathbf{v} \times \mathbf{B})$ is the electric field, $e$ is the internal
energy, and $I$ is the identity tensor.
We assume an ideal gas, i.e., $\rho e = P/(\gamma - 1)$ to close
the above equations. Here, $P = n\,k_{\rm B}\, T$, where $n = \rho / (\mu
m_{\rm p})$ is the number density, $k_{\rm B}$ is the Boltzmann's constant,
$T$ is the temperature, and $\mu m_{\rm p}$ is the mean mass of the particle.
Since we assume a fully ionised proton-electron plasma for the stellar wind, $\mu = 1/2$. 
The sound speed is $c_{\rm sw} = (\gamma P / \rho)^{1/2}$, where $P$ is the total 
electron + proton pressure and $\gamma = 5/3$ is the adiabatic index. 

We integrate the conservative forms of the equations using a Harten, Lax, Van
Leer approximate Riemann solver (\texttt{hll}) associated with a diffusive
limiter (\texttt{minmod}). The initial magnetic fields are divergenceless,
and we maintain this condition throughout the simulation, using a mixed
hyperbolic/parabolic divergence cleaning technique \citep{Dedner2002}.

The typical setup used in our simulations is as follows.  We use a grid of
128 cells in the radial direction, 48 in the polar angle direction, $\theta
\in [0, \pi]$,   and 96 in the azimuthal angle direction, $\phi \in [0,
2\pi]$. All cells are equidistant in the radial direction.  We use a
characteristic length in our simulations equal to the size of the exoplanet
Proxima b, $L = R_{\rm p} = 7\times10^8$ cm,  and a characteristic stellar wind velocity of
$V = 10^7$ cm\,s$^{-1}$, respectively. The effective numerical magnetic
Reynolds number ($R_{m}=V L/\eta$, where $\eta$ is the numerical magnetic
diffusivity) and the kinetic Reynolds number ($R_{e}=V L/\nu$, where $\nu$ is
the numerical viscosity) have values of around $1000$.  
We did not include an explicit value of the dissipation in the model, hence the numerical magnetic diffusivity regulates the typical reconnection in the slow (Sweet–Parker model) regime. A detailed discussion of the numerical magnetic and kinetic diffusivity of the model is provided in 
\cite{Varela2018}.

Our computational domain consists of a thick spherical shell centered around the exoplanet, with
the inner boundary set at \Rin\ = 2.2\,\Rp, where \Rp\ is the radius of the
planet Proxima b, and the outer boundary \Rout\ = 30\,\Rp.  The upper
ionosphere model extends between the inner boundary, \Rin, and $R$ =
2.5\,\Rp. The upper ionosphere model is based on the electric field generated by the field-aligned currents providing the plasma velocity at the upper ionosphere, and we describe the model in detail
in Appendix \ref{app:upper-ionosphere}.  We set a cutoff radius for \Bimf\ of $R_{\rm c} = 6\,$\Rp. This initial value corresponds to the approximate magnetopause standoff distance for the space weather conditions analysed in this study, and represents the region where the magnetic field of Proxima b is stronger than the interplanetary magnetic field.

We divided the outer boundary in two regions: an upstream region, where we fixed the stellar wind parameters, and a downstream region, where we assumed the null derivative condition ($\frac{\partial}{\partial r} = 0$) for all fields. Regarding the initial conditions of the simulations, we defined a paraboloid with the vertex at the dayside of the planet  as $x <
A - (y^2 + z^{2} / B)$, with $(x,y,z)$ the Cartesian coordinates, $A = R_{c}$
and $B = R_{c}*\sqrt{R_{c}}$ where the velocity is null. 
We adjusted  the
density profile to keep the Alfv\'{e}n velocity, $\mathrm{v}_{A} = B
/ \sqrt{\mu_{0}\, \rho_{\rm sw}}$, constant. Here, $\rho_{\rm sw} = n_{\rm sw}\, \mu \, \mathrm{m_p}$ is the mass density, \nsw\ is  the particle number, and  $\mathrm{m_p}$ the proton mass. In practice, since some of the simulations had a large interplanetary magnetic field, we used values of $\mathrm{v_A}$  in the range 
 $(2.6 - 5.0) \cdot 10^{4}$ km/s (a fixed value for each simulation). We note that this Alfvén speed upper limit is defined in the PLUTO simulations to control the time step of the simulations. Namely,  the Alfvén speed value sets the Alfvén time of the simulation, which is linked to
the simulation time step. The larger $\mathrm{v_A}$, the smaller is the time step, which may render the simulations extremely costly, computationally speaking. We emphasise that this condition only applies to the upper ionosphere domain.

\begin{figure}[h]
\centering
\resizebox{\hsize}{!}{\includegraphics[width=\columnwidth]{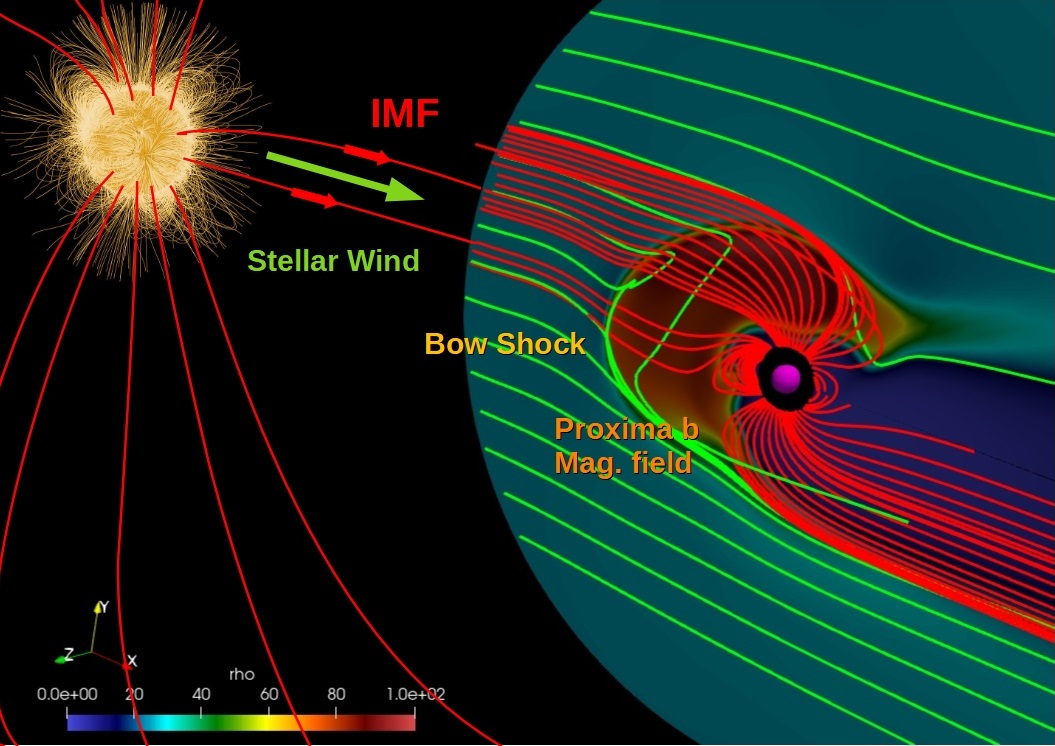}}
\caption{
Sketch of the magnetospheric interaction in the star-planet system Proxima - Proxima b. 
The stellar wind velocity and the interplanetary magnetic field (IMF) 
streamlines (assumed to be radial--see text in Sect.~\ref{sec:simulations}) are drawn in green and red, respectively. Note that there are magnetic field lines from the star (yellow circle) that are connected to the exoplanet Proxima b (purple circle).  The density distribution is shown as a color scale, normalized to the value of the stellar wind density, $\rho_{\rm sw} = \mu m_{\rm p} n_{\rm sw}$ (see Table~\ref{tab:pluto-params}). The sketch represents a super-Alfvénic case, where the ram pressure of the wind dominates over the stellar wind magnetic field pressure, so a bow shock is formed at the dayside of the planet. The Proxima star is beyond the limits of our simulation domain.
}
\label{fig:sketch}
\end{figure}

Our simulations use a frame where the
$z$-axis is given by the Proxima b magnetic axis pointing to the magnetic
north pole,  the star-planet line is located in the $XZ$ plane (with $x_{\rm
star} > 0$; stellar magnetic coordinates) and the $y$-axis completed a right
handed system.
We show in Fig.~\ref{fig:sketch} a sketch of the overall geometry for our
numerical simulations, including the M dwarf star Proxima, its planet Proxima
b, and the planetary and interplanetary magnetic field lines.  We rotate the
axis of the (dipolar) magnetic field of the planet Proxima b by $90^{\degree}$ in
the YZ plane with respect to the grid poles to avoid numerical issues (no
special treatment was necessary for the singularity at the magnetic poles).
For simplicity, we consider that the (unknown) rotation and magnetic axes of Proxima b coincide. We assume that the rotation axis makes an angle of $23.5^{\degree}$--as for the Earth--with the normal to the orbital plane (the ecliptic). To simulate different inclinations $i$, we modify the orientation of the
interplanetary magnetic field (IMF) and the stellar wind velocity vectors.

Our model reproduces the relevant global
magnetosphere structures, such as the magnetosheath and magnetopause, as demonstrated for the case of the Hermean magnetosphere
\citep{Varela2015,Varela2016v120,Varela2016v122}, although we acknowledge that it  does not resolve the plasma depletion layer as a decoupled
global structure from the magnetosheath, since the model lacks the required resolution.

We note that the reconnection between the interplanetary and Earth magnetic
field is instantaneous--no magnetic pile-up on the planet dayside-- and stronger (enhanced erosion of the planetary magnetic field) because the magnetic diffusion
of the model is stronger than in the real plasma. Nevertheless, the effects of the
reconnection region on the depletion of the magnetosheath and the injection of
plasma into the inner magnetosphere are correctly reproduced to a first
approximation. Finally, we note that we do not include the planet rotation and orbital motion in the
current model yet either, and we leave this for future work.

The magnetosphere response to the stellar wind and the IMF shows several interlinked phases that must be distinguished. First, the response of the dayside magnetopause and magnetosheath can affect the magnetosphere standoff distance, the plasma flows towards the inner magnetosphere, and/or the location of the reconnection regions, among other things. Next, the response of the magnetotail, which is followed by the ionospheric response and, subsequently, by the ring current response. In this paper, our analysis is mainly focused on the dayside response of the magnetosphere; we also discuss some implications regarding the magnetic field at the nightside, although we do not aim at a detailed analysis of the magnetotail. Finally, the response of the ionosphere and ring current are beyond the scope of the present study.

We assume that a simulation is completed when it reaches a steady state. Therefore, we do not include dynamic events caused by the evolving space weather conditions (we do not modify the stellar wind and IMF parameters along the simulation). Typically, a simulation reached steady state after $\tau = L / V = 15$ code times, equivalent to $t \approx 16$ min of physical time, although the magnetosphere topology on Proxima b dayside is steady after $t \approx 11$ min, or about 10 code times. Consequently, the code can accurately reproduce the magnetosphere response if space weather conditions are roughly steady for time periods of $t = 10-15$ min.

\section{Proxima b as a case study}
\label{sec:simulations}

Previous 3D MHD studies of the space weather of Proxima b showed that the planet may be subject to stellar wind pressures of up to three orders of magnitude higher than those experience by the Earth from the solar wind (e.g. \citealt{Garraffo2016}). Those authors found that Proxima is also subject  to pressure changes of 1$-$3 orders of magnitudes within a day, dramatically altering the magnetopause standoff distance of the planet by factors of 2 to 5. \citet{Garraffo2016} also found that Proxima b likely passes in and out fo the Alfvén surface, thus exposing the planet to both subsonic and supersonic wind conditions.
More recent MHD simulations of the space weather of Proxima b by \citet{Garraffo2022} confirmed the finds described above, and showed that the large-scale magnetic field of the star does play a very significant influence on the ambient stellar wind, while the small-scale field does not.  
\citet{Kavanagh2021} also carried out MHD simulations of Proxima Cen, using an Alfvén wave-driven stellar wind model. They found that the
mass loss rate of Proxima Cen stellar wind was \Mdotstar $\simeq$0.25 \Mdotsun, corresponding to \nsw $\sim 600$ cm$^{-3}$. 
For comparison, the simulations carried out by \citet{Garraffo2016}
imply \Mdotstar $\simeq$0.75 \Mdotsun\
(\nsw $\sim 1800$ cm$^{-3}$).  
\citet{Kavanagh2021} found, contrary to \citet{Garraffo2016}, that the orbit of Proxima b lied always beyond the Alfvén surface. In this case, there is no sub-Alfvénic star-planet interaction,  and therefore no radio emission from such (sub-Alfvénic) interaction is expected to arise.

Here, we apply the 3D MHD PLUTO code with two main goals. First, to study and discuss the habitability of Proxima b, 
focusing on the effects of the intensity of the planetary magnetic field,
\Bp, and its inclination, $i$.  
As a direct estimator of planet habitability, we determine for each simulation run the standoff magnetopause distance, \Rmp. Our second main goal is
to determine the expected radio emission arising from the dissipated power at the reconnection sites on the magnetosphere, which we directly measure from our simulations. This is relevant for prospects of using radio observations to detect exoplanets and, even more important, to directly measure their respective magnetic fields.  

Our simulations consider both calm and extreme, i.e. CME-like, space weather conditions (see Table~\ref{tab:pluto-params}). In turn, for calm space weather conditions, we run simulations for two broadly different scenarios: 
(1) the planet is in the sub-Alfvénic regime, where the stellar wind
magnetic pressure dominates over the wind's ram pressure; and (2) the planet
is in the super-Alfvénic regime, where the ram pressure of the wind is
larger than the magnetic pressure and a bow shock is formed.   The distinction between the sub- and super-Alfvénic regimes can be easily parameterized by the Alfvén Mach number, $M_A = v_{\rm sw}/v_A$, where $v_{\rm sw}$ is the speed of the plasma relative to the planetary body, and $v_A$ is the Alfvén speed. $M_A$ < 1 corresponds to the sub-Alfvénic regime, while $M_A$ > 1 corresponds to the super-Alfvénic one. The IMF is purely radial in the model, and assumes an intense stretching effect of the stellar wind on the magnetic field lines of Proxima. This leads to a dominant radial component of the IMF at the orbital distance of Proxima b.

\begin{table*}[htb]
  \centering
  \caption{\label{tab:pluto-params} \texttable{Main parameters of the PLUTO simulations} }
  \begin{tabular}{lccc} \toprule 
      \multirow{3}{*}{Parameter} & 
      \multicolumn{2}{c}{Calm space weather} & 
       Extreme space weather \\
      & {Sub-Alfv\'enic} & {Super-Alfv\'enic} & {CME} \\ 
      & {scenario} & {scenario} & {scenario} \\ 
      \toprule \toprule   
      $n_{\rm sw}$ [cm$^{-3}$]             & 50   & 50  & 250  \\
      $|v_{\rm sw}|$ [$10^7$ cm\,s$^{-1}$] & 5    & 10  & 25 \\
      $B_{\rm IMF}$ [mG]                   & 3.2  & 1.6 & 16 \\
      \midrule
      $B_{\rm Proxima}$ [G]                & 1200 & 600  & 600\\
      $B_{\rm Proxima\, b}$ [G]            & 0.16$-$1.28 & 0.16$-$1.28 & 0.16$-$0.64 \\  
      $i$                                  & [0$^{\degree}$ - 90$^{\degree}$] & [0$^{\degree}$ - 90$^{\degree}$]  & [0$^{\degree}$ - 45$^{\degree}$] \\
      \bottomrule
  \end{tabular}
\end{table*}

We show in Table~\ref{tab:pluto-params} the values for the most relevant
parameters of our PLUTO simulations and, for comparison, in Table~\ref{tab:params-comparison} we show the parameters used in the MHD simulations by \cite{Garraffo2016} and \citet{Kavanagh2021}.

We note that none of those works discuss the radio emission generated in the magnetosphere of Proxima b, which is one of the main goals of the present study.

The stellar wind density values, \nsw, in our simulations are 50$-$250 times  larger than
  in the solar wind around the Earth, and correspond to mass-loss rates of the stellar wind,  
  \Mdotstar, of $\simeq$ 0.02 \Mdotsun\ 
and $\simeq$ 0.1 \Mdotsun\  for the calm and CME-like scenarios, respectively. Those values are in agreement with the upper
  limit for the stellar wind of Proxima Cen of 0.2 \Mdotsun\ from astrospheric
absorption measurements, found by \citet{Wood2001}.

In the sub-Alfvénic case, we use a value of 3.2 mG for \Bimf\, which corresponds to the
extrapolation of the Proxima Cen magnetic field at the pole (1200 G) at the
orbital distance of Proxima. In the super-Alfvénic case, we set a value of \Bimf\ half as large as in the sub-Alfvénic scenario. 
However, we note that the magnetic field value of Proxima could be lower (e.g. \citealt{Klein2021}). 
Still, the values of Proxima Cen magnetic field considered here are consistent with previous works. Indeed,
\citet{Reiners2008} found a value of 600$\pm$150 G for the average surface magnetic field. Also, 3D MHD simulations by \citet{Yadav2016} indicate that Proxima Cen undergoes strong variations of its
magnetic field surface intensity during its $\sim$7-yr long cycle, starting from
about 500 G and going up to 2000 G. 

For the CME-like scenario, we set values of the stellar wind density and speed, and of \Bimf\ 5, 2.5 and 10 times larger than in the standard Super-Alfvénic scenario. 

In our simulations, and
without any loss of generality, we fixed the stellar wind
temperature to $T_{\rm sw} = 3 \times 10^5$ K and the planet surface
temperature $T_{\rm planet} = 1000$ K. We also note that we assume a dipolar
magnetic field for the exoplanet.

\begin{figure}[htb!]
\centering
\resizebox{\hsize}{!}{\includegraphics[width=\columnwidth]{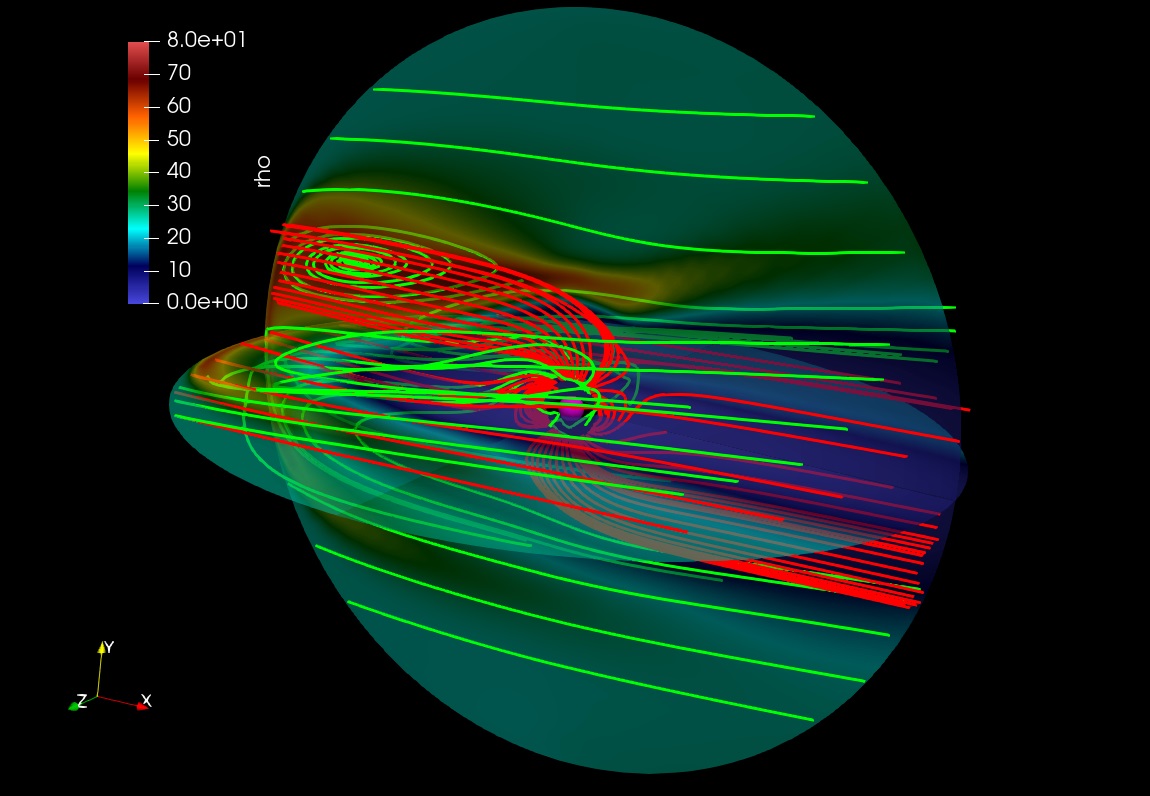}} \\
\resizebox{\hsize}{!}{\includegraphics[width=\columnwidth]{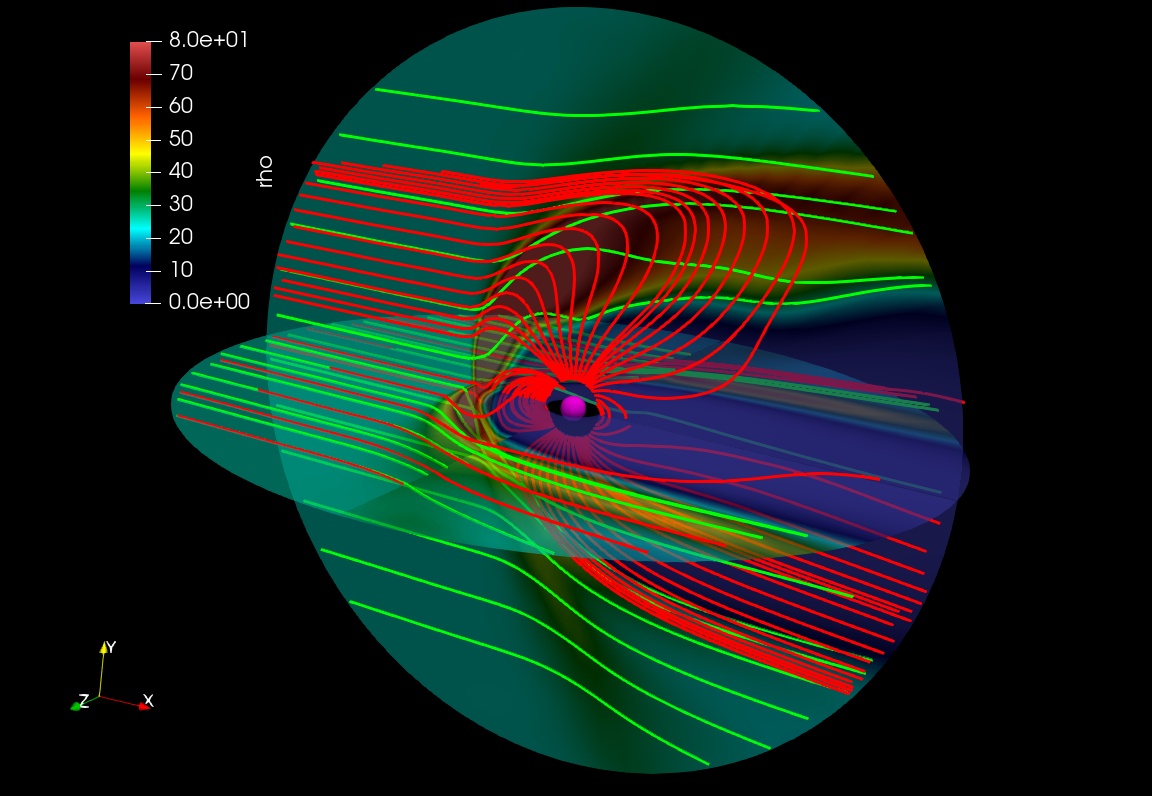}}
\caption{
Plots of density distribution (coloured scale, in particles cm$^{-3}$) for two sub-Alfvénic (top) and super-Alfvénic (bottom) cases of our PLUTO simulations. Green lines correspond to streamlines of the stellar wind velocity, and red lines to magnetic field lines of both the stellar wind and the planet. In both cases, $B_{\rm p} = 0.32$ G and the tilt of the planet is $i = 0$ deg, i.e., $B_{\rm p}$ is perpendicular to $B_{\rm imf}$).
In the sub-Alfvénic case, no bow shock is formed, and the resulting radio emission originates only from the reconnecting region of the exoplanet magnetosphere.
In the super-Alfvénic case, a bow shock is formed. The radio emission originates both from the reconnecting region of the magnetosphere and from the bow shock (see Sect.~\ref{sec:radio}). [Movies of those simulations, from beginning to end, when a steady solution is reached, can be downloaded from here.] 
}
\label{fig:alfven}
\end{figure}


\begin{figure}[htb!]
\centering
\resizebox{\hsize}{!}{\includegraphics[width=\columnwidth]{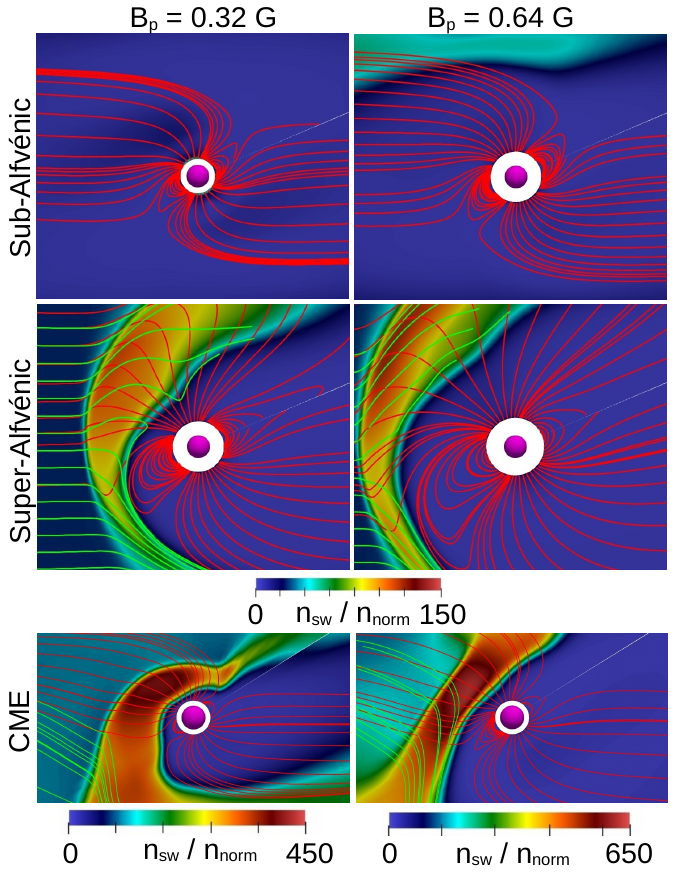}}
\caption{
Plots of density distribution (coloured scale, in particles cm$^{-3}$) for simulations under calm space weather conditions (top and middle panels), and under a CME (bottom panels), for two different magnetic field intensity values of the exoplanet Proxima b. Green and red lines are as described in Fig.~\ref{fig:alfven}. 
} 
\label{fig:Super-Sub-CME}
\end{figure}

\begin{table}[htb]
  \centering
  \caption{\label{tab:params-comparison} \texttable{Parameters in other 3D MHD simulations} }
  \begin{tabular}{lll} \toprule 
      Parameter & Garraffo+2016 & Kavanagh+2021\\
      \toprule
      \Mdotstar [\Mdotsun]                 & 0.75   & 0.25   \\
      $n_{\rm sw}$ [cm$^{-3}$]             & 1800   & 600   \\
      $|v_{\rm sw}|$ [$10^7$ cm\,s$^{-1}$] & 13, 16 & 3$-$12 \\
      $B_{\rm Proxima}$ [G]                & 600, 1200 & 200\\
      $B_{\rm Proxima\, b}$ [G]            & 0.1, 0.3 &  Unmagnetized \\
      $i$                                 & 10$^{\degree}$, 60$^{\degree}$ & 0$^{\degree}$ \\    
      \bottomrule
  \end{tabular}
\end{table}

For each sub- and super-Alfvénic case, we used the following values of the
intrinsic planetary magnetic field: 0.16, 0.32, 0.64 and 1.28 G, with the
value of 0.32 G being the nominal average magnetic field for the Earth. For
the study of the effect of the tilt, we fixed the magnetic field to be \Bp\ =
0.32 G, i.e., an Earth-like magnetic field, and used the following tilt values: $i
\in$ [0, 7, 15, 23.5, 30, 45, 60, 75, 83, 90] deg.  
A value of $i=0$ deg
corresponds to the case when \Bp\ is perpendicular to \Bimf,  and $i=90$ deg
corresponds to the case when \Bp\ is 
anti-aligned with respect to \Bimf.
For the computation of the radio emission in Sect. \ref{sec:radio}, we fixed the
efficiency factor in converting Poynting flux to radio emission to a value of
$\beta = 2\cdot10^{-3}$ \citep{Zarka2007,Zarka6}. 

In Fig.~\ref{fig:alfven} we show the final, steady situations of two
representative cases of our simulations.  The top panel corresponds to a
sub-Alfvénic scenario ($M_A < 1$). In this case, the bow shock around the exoplanet (the region with the highest density, in red) dissipates all the way throughout the simulation
and, in practice, is never formed.  We can
still see some features reminiscent of a dying bow shock and, but those would
have disappeared completely if we had extended the simulations for a longer
time. The resulting radio emission originates therefore only from the
reconnecting region of the exoplanet magnetosphere.  The bottom panel
illustrates a super-Alfvén case, when a bow shock is formed. 
The accelerated electrons producing the radio emission originate both from
the reconnecting region of the exoplanet magnetosphere and from the bow shock.

In Fig.~\ref{fig:Super-Sub-CME} we show close-up images of the region close to the planet, for the steady situation obtained from simulations under calm space weather conditions (top and middle panels), and under extreme space weather (bottom panels), for two different values of the intensity of the magnetic field of Proxima b, \Bp = 0.32 G and \Bp = 0.64 G. As shown in Fig.~\ref{fig:alfven}, when the conditions are sub-Alfvénic, no bow shock is formed, while the bow shock can be clearly seen in the rest super-Alfvénic regime. Note also that as the exoplanetary magnetic field increases, the bow shock is formed farther away and the magnetopause standoff distance is larger, thus increasing the protection of the planet against damaging particle and radiation from outside (see Sect.~\ref{sec:magnetopause} for a detailed discussion). Note also how the particle density is much larger in the bow shock region formed under extreme weather conditions (bottom panels), compared to that formed under calm space weather.

\section{Proxima b habitability - magnetopause standoff distance}
\label{sec:magnetopause}

In this section, we determine the values of the magnetopause standoff distance, \Rmp, in our simulations, both as a function of the tilt angle of Proxima b and of its magnetic field. We use \Rmp\ to infer the habitability of Proxima b.  Namely, if \Rmp\ $\leq$ \Rp, then there is direct precipitation of damaging particles, and  the habitability is strongly constrained. We note, however, that for the simulations under extreme weather conditions we set up a minimum value of $R_{in}$ = 1.5\ , \Rp\ since for smaller values there were numerical issues with the boundary conditions of the internal region. Therefore, for the CME simulations, if \Rmp\ $\leq 1.5\,R_{\rm p}$, we considered that there was direct particle precipitation on the exoplanet.

In the analysis of our simulations, we defined \Rmp\ as the radial distance to
the last closed magnetic field line on the exoplanet dayside (similar to the
case of Ganymede, see e.g. \citealt{Kivelson}), at $0^{\degree}$ longitude in
the ecliptic plane (see Fig.~\ref{fig:Rmp}). 
We therefore directly measured \Rmp\ from our simulations, instead of using the (theoretical) magnetopause standoff distance, \Rmp\ , which can be obtained from the
balance between the pressures of the stellar wind and that of the exoplanet magnetosphere. 
The pressure of the stellar wind includes the 
dynamic, thermal and magnetic pressure components; the pressure of the exoplanet 
magnetosphere includes the thermal
and (dipolar) magnetic field pressure. 
From that pressure balance, one gets $R_{\rm mp}/R_{\rm p}$
(e.g., \citealt{Varela2022SpWea}): 

\begin{figure}[htb!]
\centering
\includegraphics[width=\columnwidth]{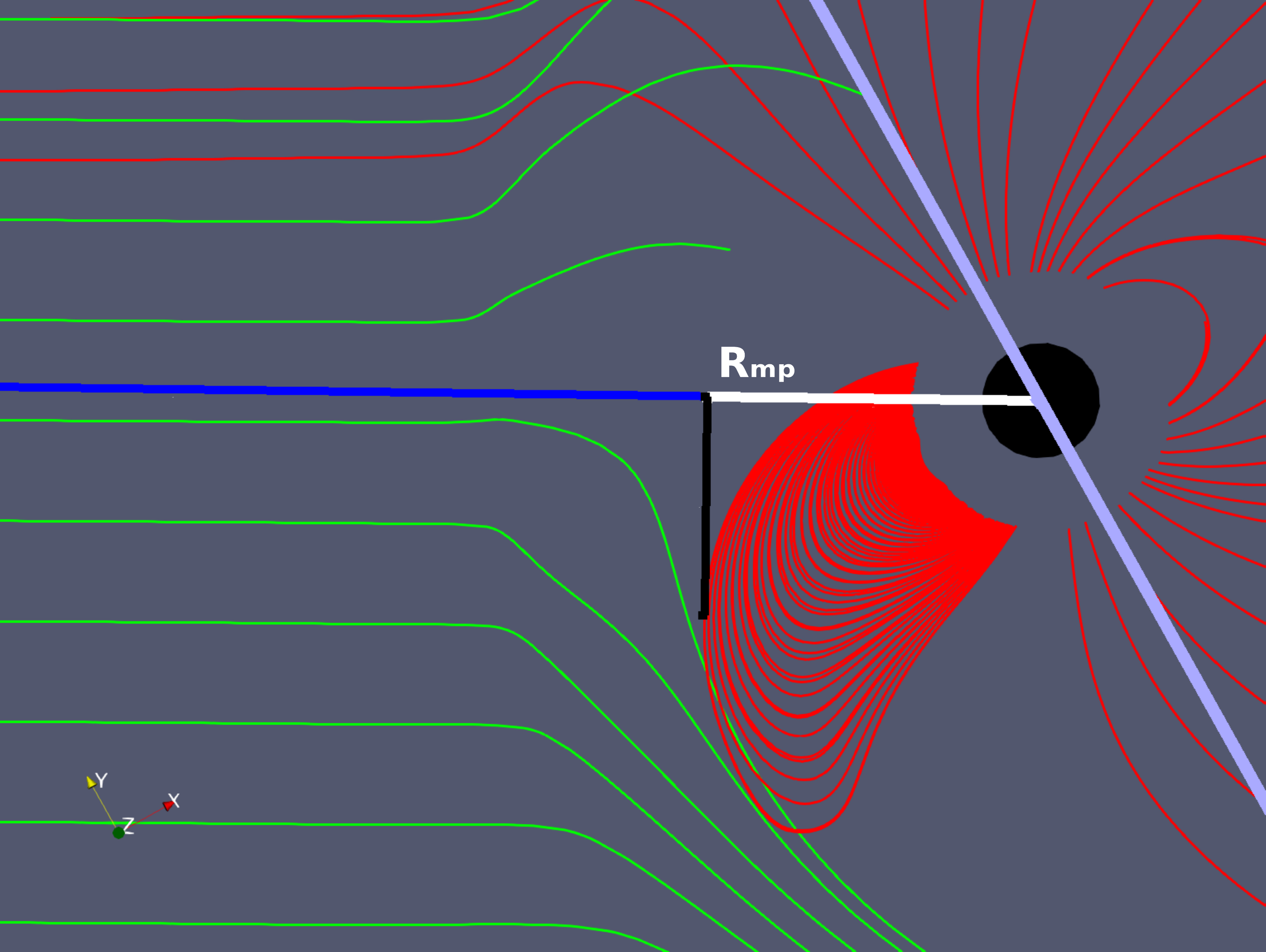}
\caption{\label{fig:Rmp}
Illustration of the measurement of the magnetopause standoff distance in our simulations.
The black circle corresponds to the planet.
The standoff distance is the radial distance (white line) to the projection of the last closed magnetic field of the planet (in red) on the line connecting the star and the planet (in blue). 
The simulation depicted here corresponds to a super-Alfv\'enic case, where the inclination of the magnetic axis of the planet with respect to the ecliptic is $30^{\degree}$ (purple line)  and $B_{\rm p} = 0.32$ G.  
} 
\end{figure}

\begin{equation}
\label{eq:Rmp}
\frac{R_{\rm mp}}{R_{\rm p}} = 
        \left[ \frac{\alpha \mathcal{M_{\rm p}}^{2} / \pi}{ m_{\rm p} n_{\rm sw}  v_{\rm sw}^{2} 
        + \frac{B_{\rm IMF}^{2}}{4\pi} 
        + \frac{2 m_{p} n_{\rm sw}  c_{\rm sw}^{2}}{\gamma} 
        - m_{\rm p} n_{\rm bs} v_{\rm th, msp}^{2}} \right]^{1/6},
\end{equation}

\noindent 
where 
$\mathcal{M}_{\rm p}$ is the exoplanet dipole magnetic field moment, $\alpha$  the dipole compression coefficient ($\alpha \approx 2$, 
\citealt{Gombosi1994}), $n_{\rm bs}$  the particle number density in the bow shock, and $v_{\rm th, msp}$ the speed of the thermal electrons in the magnetosphere, and the above expression is in cgs units. If $R_{\rm mp}/R_{\rm p} \leq 1$, then there is direct
precipitation of the stellar wind plasma particles towards the exoplanet
surface, and habitability is not possible.

We emphasise that Eq.~\ref{eq:Rmp} above is just an approximation, and the actual value of \Rmp\ can depart significantly from the true value, as the above expression does not take into account a number of important effects. First, there is no topological consideration at all, and all terms in the equation are treated as simple scalar values. Second, this expression does not include the effect of the
reconnections between the IMF with the exoplanetary magnetic field lines. Therefore, Eq.~\ref{eq:Rmp} assumes a compressed dipolar magnetic field, while ignores the
orientation of the IMF. Third, 
Eq.~\ref{eq:Rmp}  is only valid if
the reconnection between the IMF and Proxima b magnetic field is rather weak.

As a consequence, the application of Eq.~\ref{eq:Rmp} to calculate the magnetopause
standoff distance may depart signficantly with respect to the values determined from 3D MHD simulations, such as the ones we carry out in this work.  
In Fig. \ref{fig:standoff} we show the dependence of the normalized magnetopause standoff distance, $r = R_{\rm mp}/R_{\rm p}$, with the planet tilt (left panels) and with the magnetic field of Proxima b (right panels), for 
calm weather (both sub-Alfvénic and super-Alfvénic; top and middle panels, respectively) and extreme weather conditions (bottom panels), as described in Table~\ref{tab:pluto-params}.
Note that, for any given set of values of $n_{\rm sw}, v_{\rm sw}, B_{\rm IMF}$ and $B_{\rm p}$, Eq.~\ref{eq:Rmp} predicts a single value, whereas our simulations indicate values that range from 
$R_{\rm mp} \approx R_{\rm p}$ up to $R_{\rm mp} \gtrsim 7\,R_{\rm p}$, which clearly illustrates the point that the standard usage of Eq.~\ref{eq:Rmp} may lead to wrong results.

\subsection{Calm space weather scenario}

In this subsection, we discuss the results obtained for the sub-Alfvénic and super-Alfvénic simulations under calm space weather conditions.
For the simulation runs as a function of the tilt of the planet, we used \Bp\ =0.32 G (i.e., an Earth-like magnetic field).
First, note that \Rmp\ monotonically decreases with the tilt angle, independently of the Alfvénic regime where the planet is. For a wide range of tilt values, \Rmp\ is several times the planet radius, which means that the magnetic shield is enough to counter-rest the damaging cosmic radiation. For very large tilt angles, however, \Rmp\ can be essentially equal to the planetary radius, which means that any damaging radiation goes unimpeded to the surface of the planet, and habitability will be severely threatened.

\begin{figure*}[ht!]
  \centering
  \begin{minipage}{\textwidth}
    \centering
    \includegraphics[width=.4\textwidth]{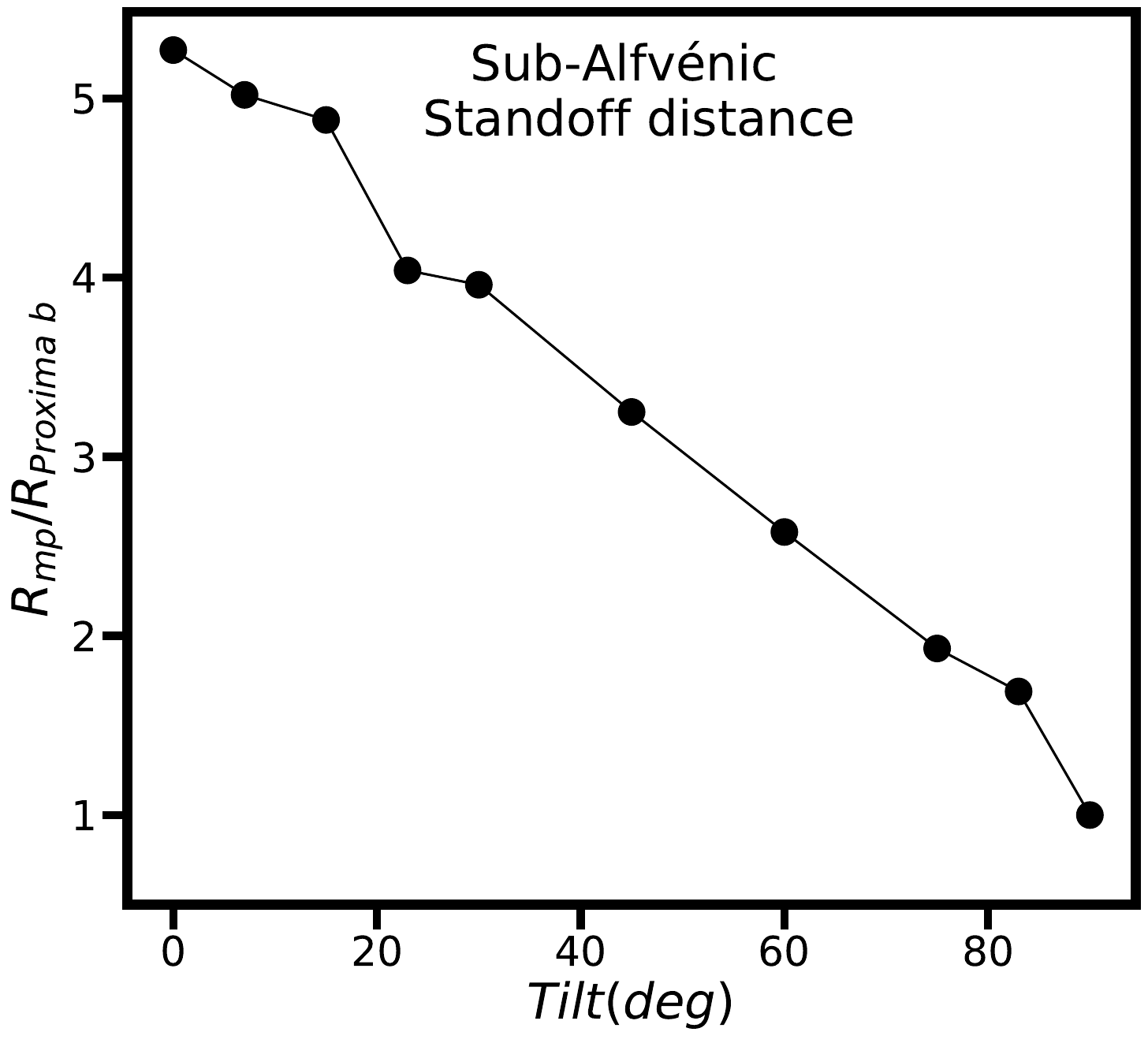}
    \includegraphics[width=.4\textwidth]{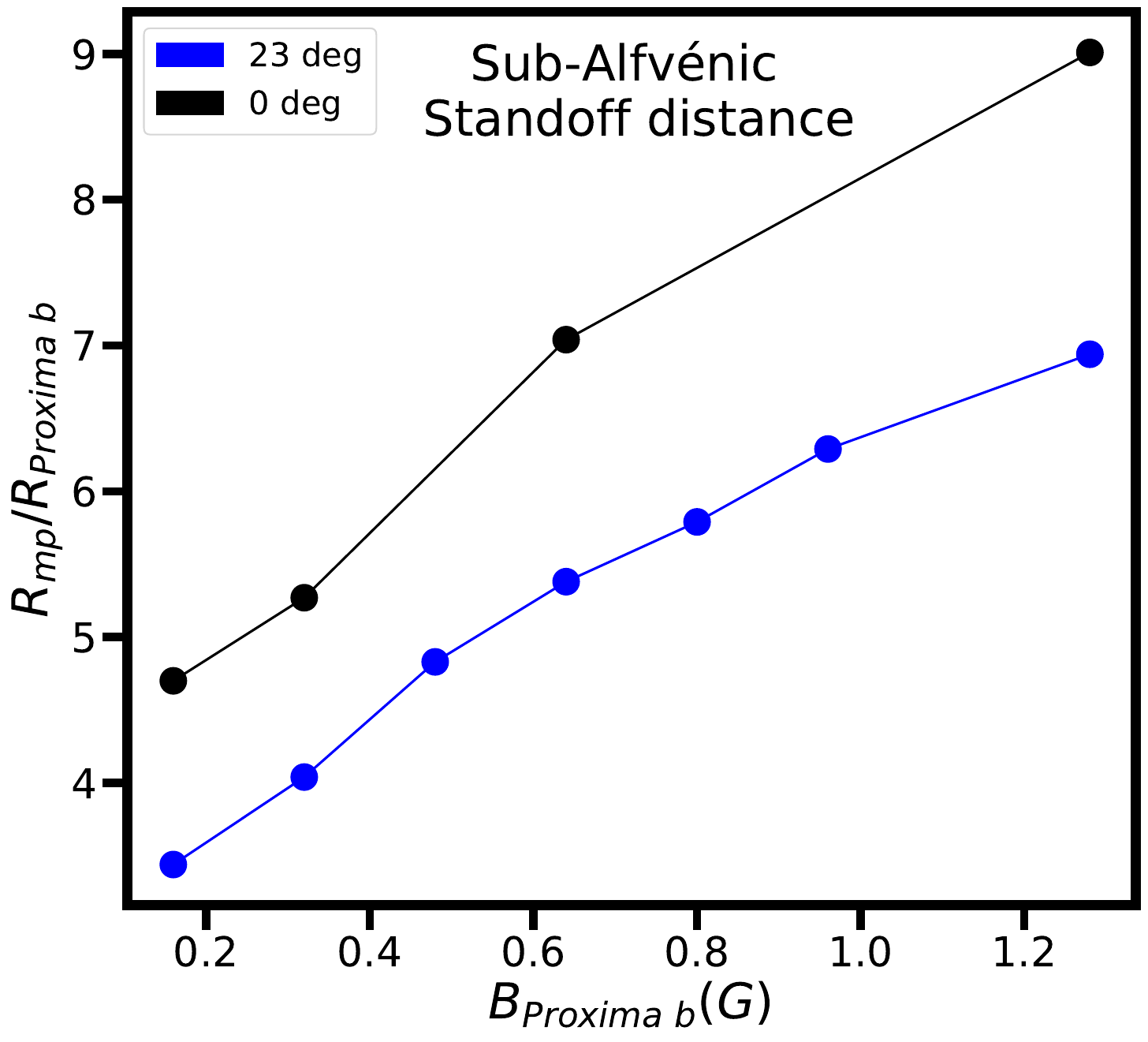}
  \end{minipage}
  \begin{minipage}{\textwidth}
    \centering
    \includegraphics[width=.4\textwidth]{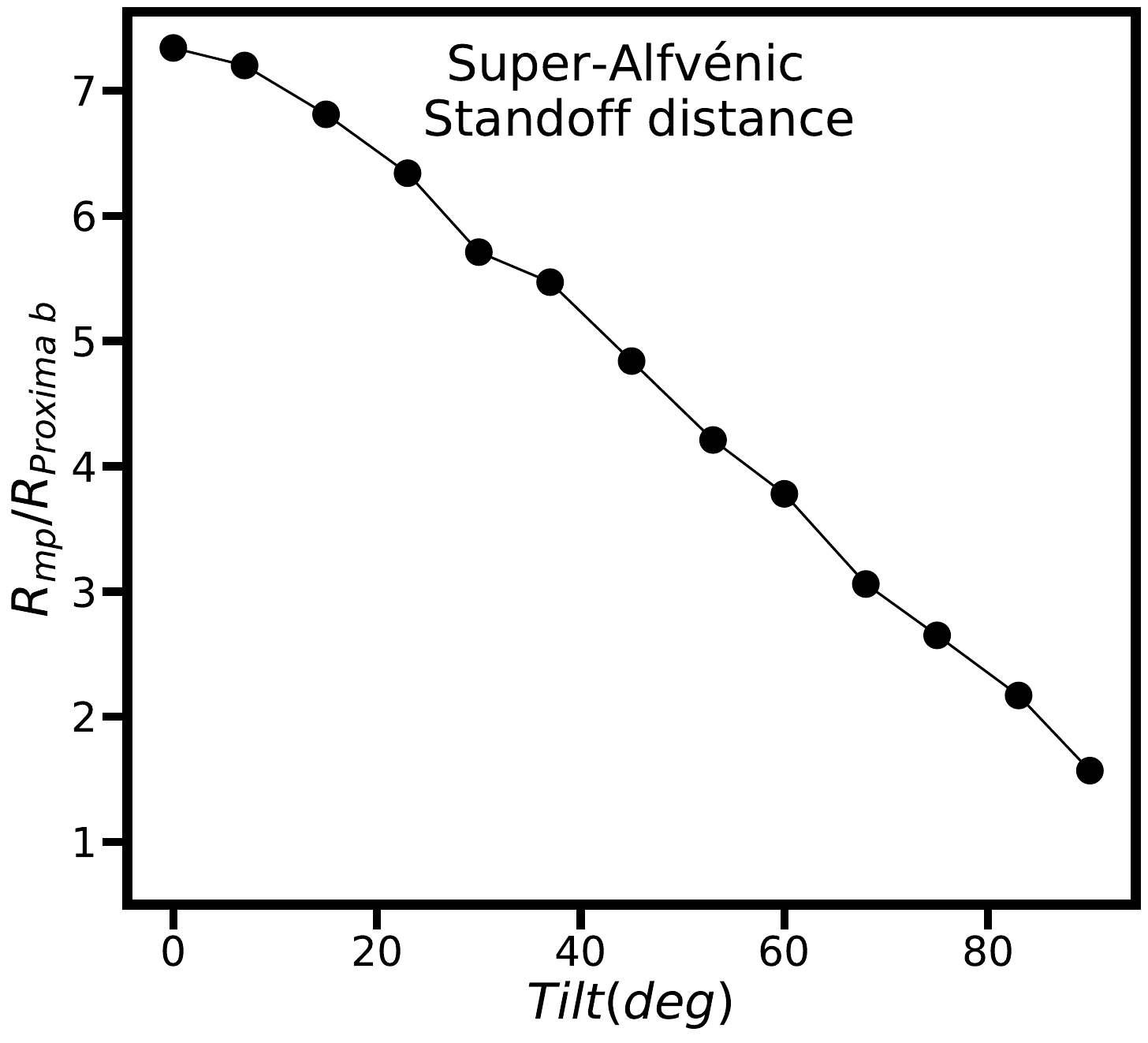}
    \includegraphics[width=.4\textwidth]{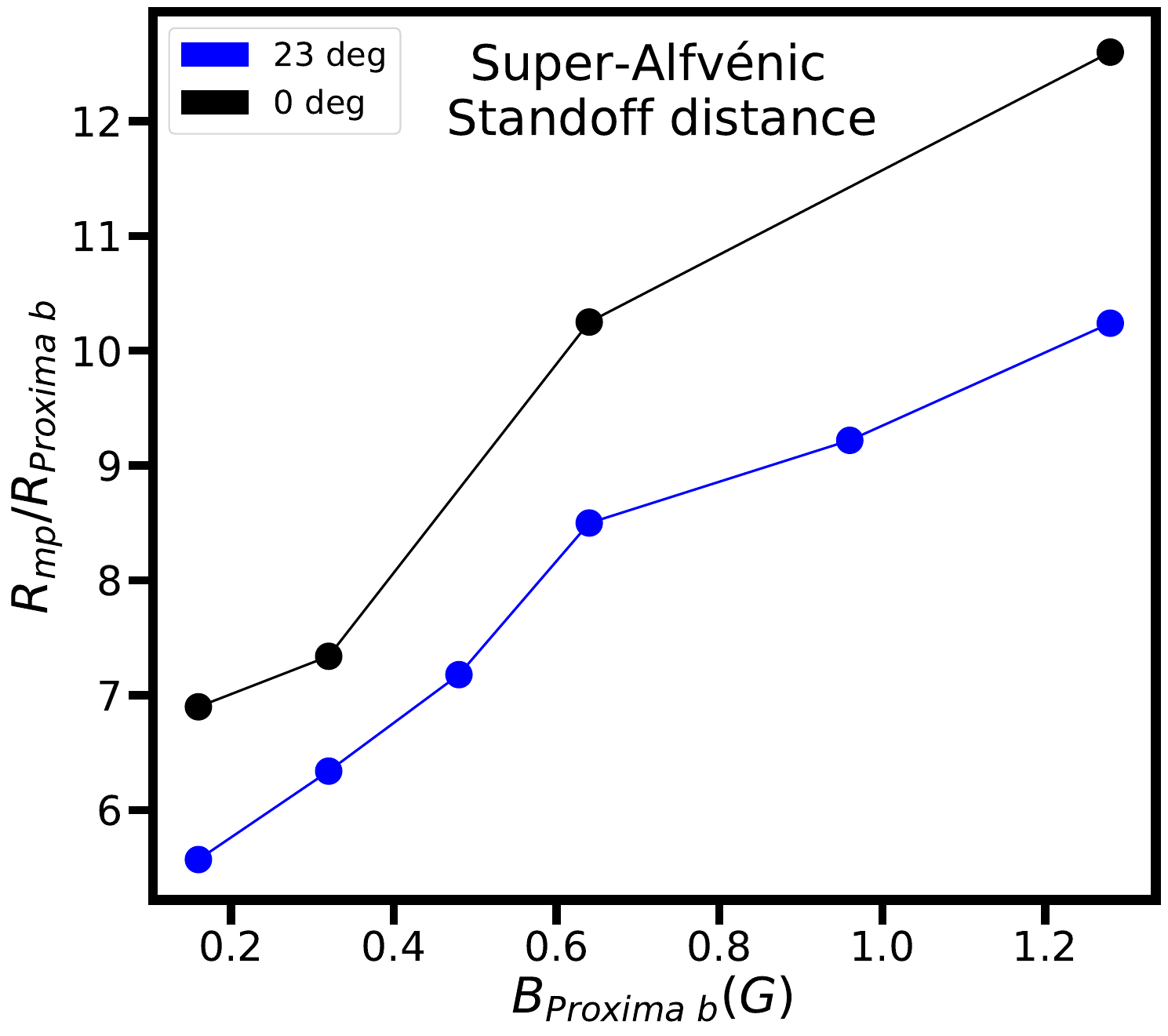}
  \end{minipage}
  \begin{minipage}{\textwidth}
    \centering
    \includegraphics[width=.41\textwidth]{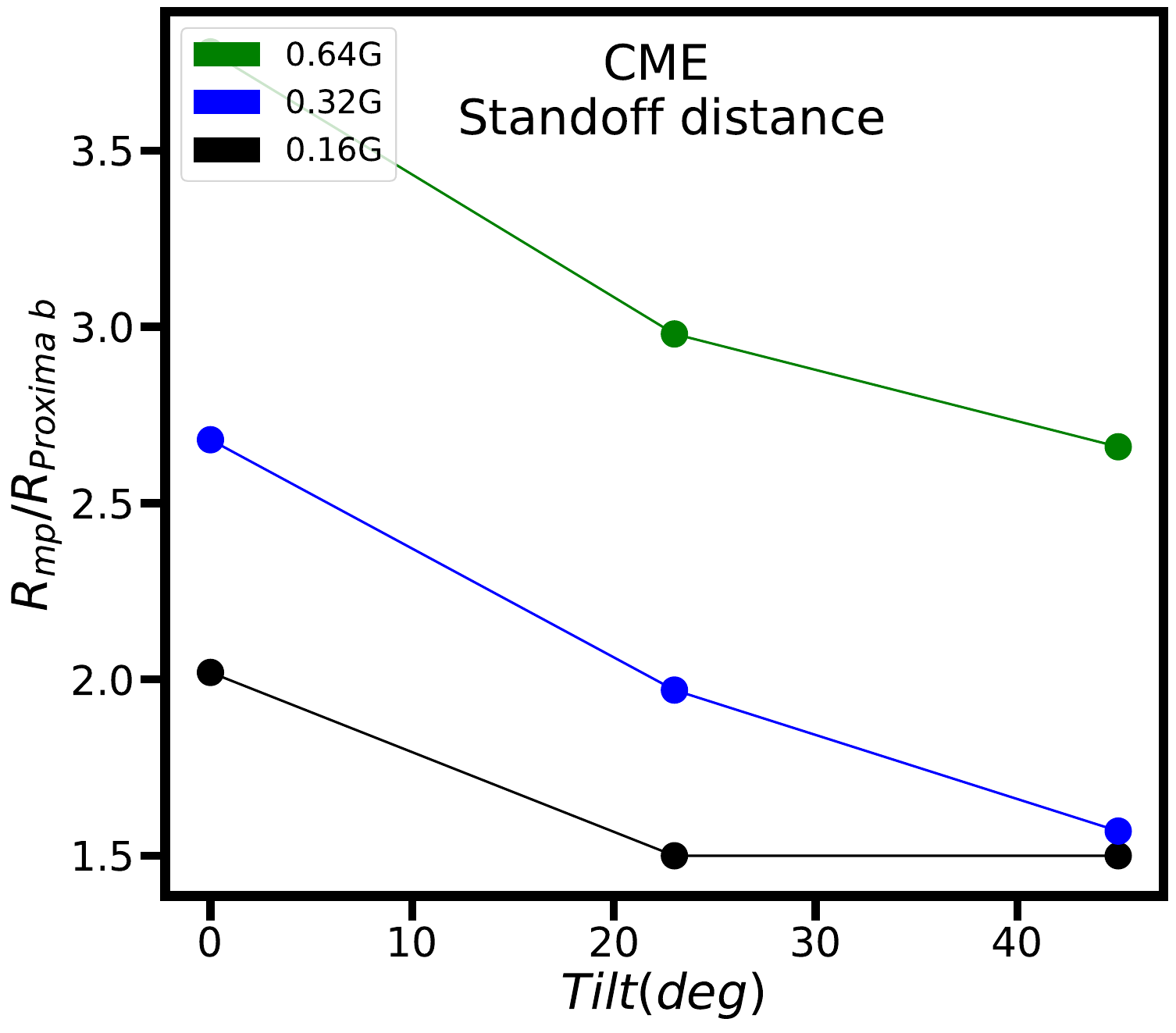}
    \includegraphics[width=.4\textwidth]{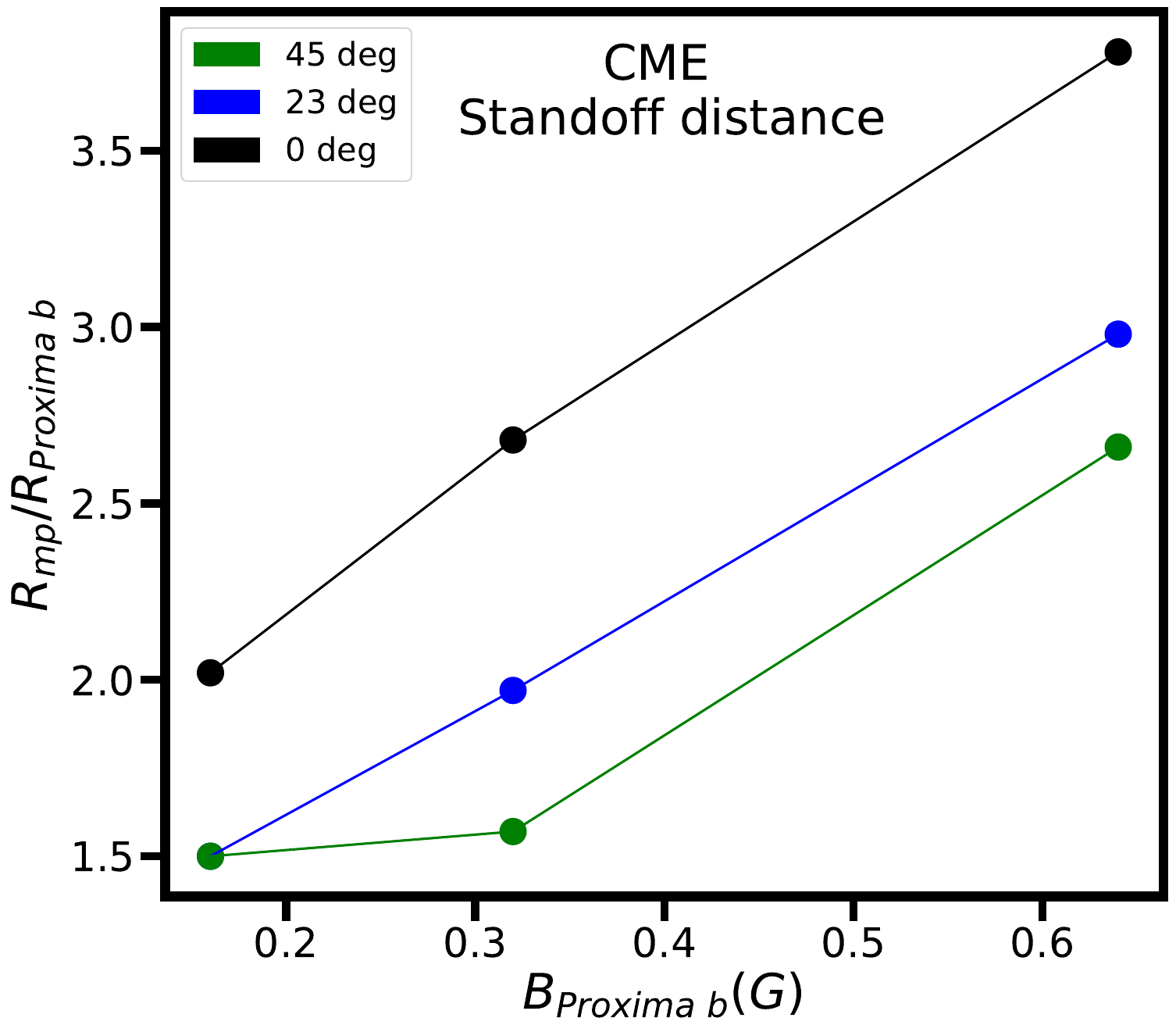}
  \end{minipage}
  \caption{Magnetopause standoff distance as a function of the tilt of the planet (left panels), and of the intrinsic planetary magnetic field, \Bp\ (right panels). Top and middle panels correspond to the sub-Alfvénic and super-Alfvénic cases, respectively, while the bottom panels correspond to a CME-like scenario. The top and middle left panels were obtained for a value of \Bp=0.32 G.
  \label{fig:standoff}}
\end{figure*}

Note that the magnetopause standoff distance is smaller in the 
sub-Alfvénic case, compared to the super-Alfvénic one. This result may seem 
counter-intuitive at first, since the dynamic pressure of the stellar wind is four times larger in the 
super-Alfvenic configuration, as the stellar wind velocity is twice as large  (see Table~\ref{tab:pluto-params}).
On the other hand, the magnetic pressure of the IMF is four times larger, so the reconnection  intensity is stronger in the sub-Alfvenic case 
(since the IMF intensity is twice as large). This  
leads to an enhanced erosion of the magnetic field of Proxima b. 
Our simulations therefore imply that in the sub-Alfvénic case the effect of the IMF erosion on the magnetic field of Proxima is stronger than the effect of the enhanced magnetosphere compression in the
super-Alfvenic case. We also recall that the pressure balance
in the sub-Alfvénic case is different from that in the super-Alfvénic case
because the effect of the thermal pressure of the bow shock disappears. 
Therefore, a direct comparison of the magnetopause standoff distance in both configurations can be misleading, since the entire pressure balance is different.    

In our study of the dependence of \Rmp\ on the Proxima b surface magnetic field, we used two different values of the tilt angle, $0^{\degree}$ and $23.5^{\degree}$, the latter case representing an Earth-like case. 
Note that, as expected, the larger the planetary magnetic field, the larger the value of \Rmp, regardless of the Alfvénic regime. More specifically,  \Rmp\ nearly doubles when \Bp\  increases from 0.16 to 1.28 G: from 4.7 $R_p$ to 9.0 $R_p$ for the sub-Alfv\'enic regime, and from 6.9 $R_p$ to 12.6 $R_p$ in the super-Alfv\'enic regime for planet with no inclination. For a planet with an inclination of $23.5^{\degree}$, as in the case of the Earth, \Rmp\ increases from 3.6 $R_p$ to 7.2 $R_p$ in the sub-Alfv\'enic regime, and from 5.6 $R_p$ to 10.2 $R_p$ in the super-Alfv\'enic regime. This trend agrees with the expected behaviour for \Rmp\, from Eq.~\ref{eq:Rmp}. Indeed, 
since \Rmp\ $\propto \mathcal{M_{\rm p}}^{1/3}$, and given that $\mathcal{M_{\rm p}} \propto B_{\rm p}$, it follows that
 \Rmp\ $\propto B_{\rm p}^{1/3}$.  Since we used values from 0.16 G up to 1.28 G,  the theoretical value of \Rmp\ increases by a factor of 8$^{1/3} = 2$, which is almost exactly the increase shown by \Rmp\  in the right panels of Fig.~\ref{fig:standoff}.  The main conclusion from those simulations is that if the magnetic field of Proxima b is Earth-like or larger, it suffices to counteract the role of the total pressure of the stellar wind.

On the other hand, the behaviour of \Rmp\ as a function of the tilt angle cannot be recovered from Eq.~\ref{eq:Rmp}. Our simulations clearly show that as the planetary tilt increases, the magnetopause standoff distance decreases, leading eventually to values that constrain the habitability of Proxima b.

We also note that for our sub-Alfv\'enic simulations we considered that the stellar wind speed was half the one we used for the super-Alfv\'enic cases, and the interplanetary magnetic field was twice as large, significantly enhancing the effect of the magnetic reconnection between the planetary and interplanetary magnetic fields.
In the sub-Alfv\'enic case magnetic erosion is therefore quite dominant, which greatly shrinks the magnetopause.
In both cases, the standoff distance is large enough that there would be no stellar wind precipitation on the planet, at least during regular space conditions. 
The exception is the case with a very high planet tilt (close to $90^{\degree}$), when the standoff distance decreases to values of 1 and 1.57 $R_p$  for the sub- and super-Alfv\'enic cases, respectively, which would make habitability not possible in the sub-Alfvénic case, as there would be direct precipitation of stellar wind particles, and in the super-Alfvénic case would be extremely vulnerable to even relatively small variations in the activity of the host star.

\subsection{CME scenario}

The bottom panels of Fig.~\ref{fig:standoff} show the results of PLUTO simulations for extreme space weather conditions around Proxima b, such as those produced by a CME from its host star. 
We recall that in those simulations we had to set up a minimum value of 1.5\,\Rp\ for the inner boundary radius of the planet, which effectively means that if the value of \Rmp\ drops down to 1.5\,\Rp, there is direct precipitation of particles onto the planet.
Since the dynamic and magnetic pressure of the wind are so large, the magnetopause standoff distance is significantly smaller with respect to either the sub-Alfvénic or super-Alfvénic cases, as a function of either the tilt angle, $i$, or of \Bp.
The bottom left panel shows, as expected, that the larger the tilt, the smaller the value of  \Rmp. In fact, for tilts above $\sim 45 \degree$, all simulations yielded values of \Rmp $\lesssim$ 1.5\,\Rp, implying that there is direct particle precipitation onto the planet. 
If the planet has a small tilt, even a relatively small value of \Bp\ is enough to shield the planet. However, if the tilt is similar to that of the Earth, or larger, \Bp\ needs to be also similar, or higher, to that of our planet to permit magnetic shielding.

\section{Radio emission from Proxima b}
\label{sec:radio}

The detection of radio emission from the magnetosphere of an exoplanet is probably the only direct way
of determining its magnetic field. This is of huge relevance for the
understanding of the interiors of exoplanets. Once the characteristics
of the exoplanet magnetic field are inferred, 
it becomes feasible to study the
space weather conditions generated by the host star on the
exoplanet orbit. Therefore, the radio emission from the exoplanet magnetosphere and the space weather conditions to which the exoplanet is subject are tightly intertwined.   
In addition, radio emission estimates from numerical
modelling are useful to guide future observations of star-planet systems,
aimed at the detection of new worlds using radio interferometers.
In this section, we present estimates of the radio emission expected to arise from the magnetospheric reconnection regions, both under calm and extreme space weather conditions around Proxima b. The predicted radio emitted power from the interaction between the stellar wind of
Proxima and the magnetosphere of its planet Proxima b is therefore one of the
main results from our numerical modelling.  

We followed the radio-magnetic Bode’s law, and used the incident magnetized flow
power and the obstacle magnetic field intensity to determine the radio
emission as $P_{R} = \beta P_{\rm B}$, where  $\beta$ is the efficiency
of converting the dissipated power, $P_{\rm B}$, into radio emission power, and $\beta \approx (2 - 10) \cdot 10^{-3}$  \citep{Zarka2018}. 
We calculated the power dissipated in the interaction between the stellar wind and the magnetosphere
at the exoplanet dayside. Irreversible processes in the
interaction convert internal, bulk flow kinetic and magnetic energy into the
kinetic energy required to accelerate the electrons along the magnetic field
lines, leading to cyclotron-maser radiation emission  by these accelerated
electrons (see \citealt{Varela2022SpWea} for details.) 
The radio emission, using the net magnetic power
deposited on the exoplanet day side, is therefore

\begin{figure*}[htb!]
  \centering
  \begin{minipage}{\textwidth}
    \centering
    \includegraphics[width=.31\textwidth]{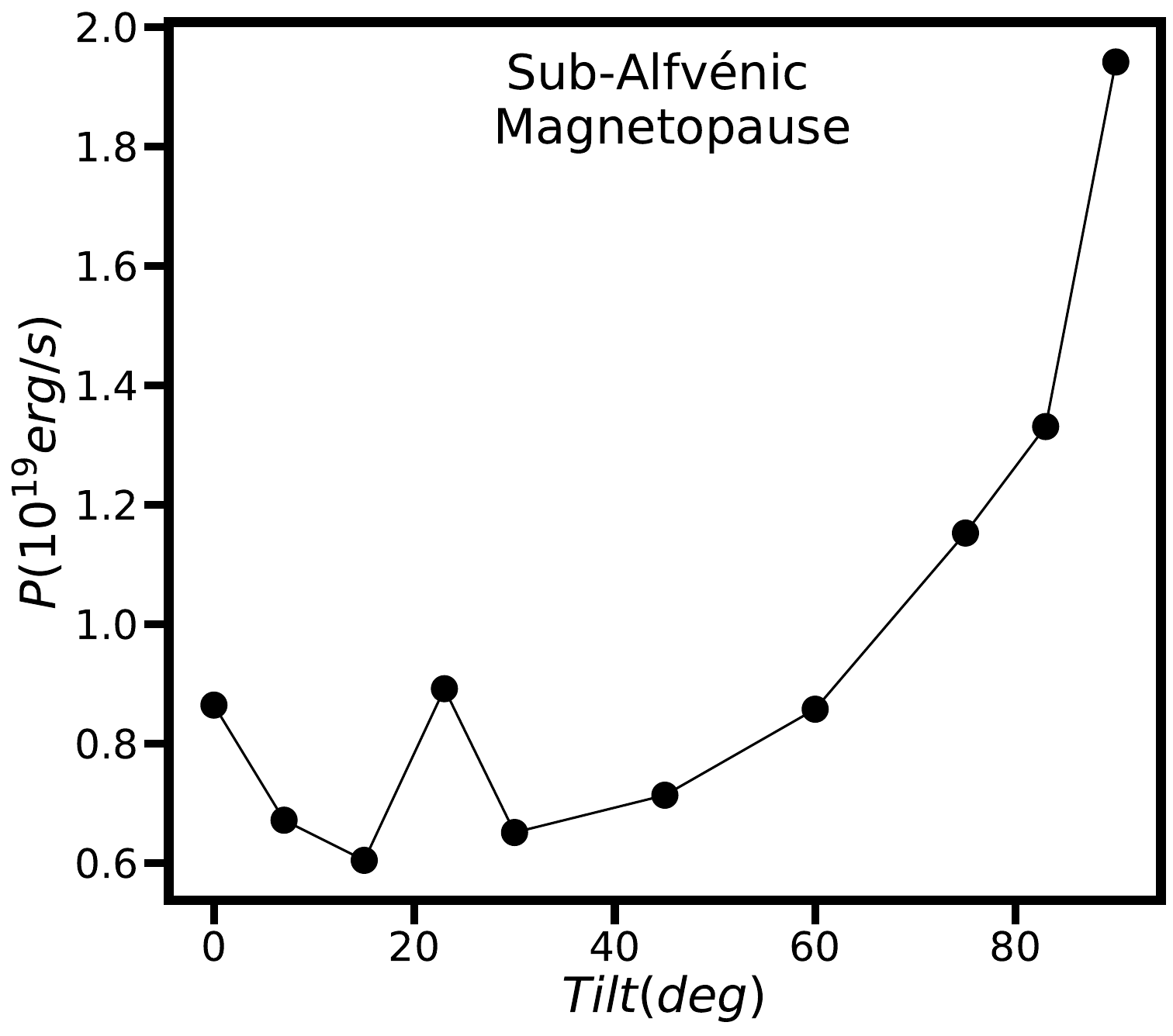}
    \includegraphics[width=.31\textwidth]{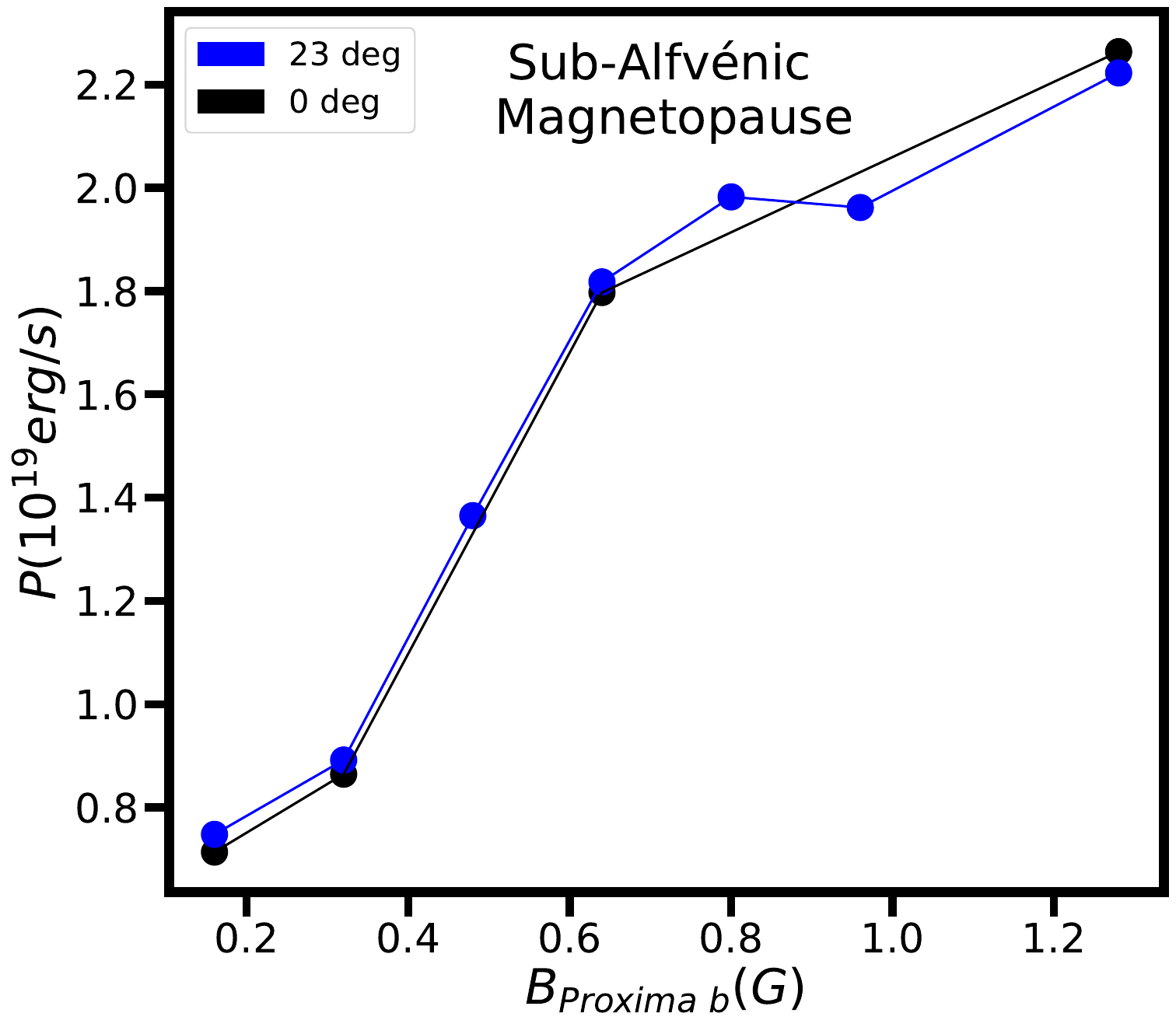}
  \end{minipage}
  \begin{minipage}{\textwidth}
    \centering
    \includegraphics[width=.31\textwidth]{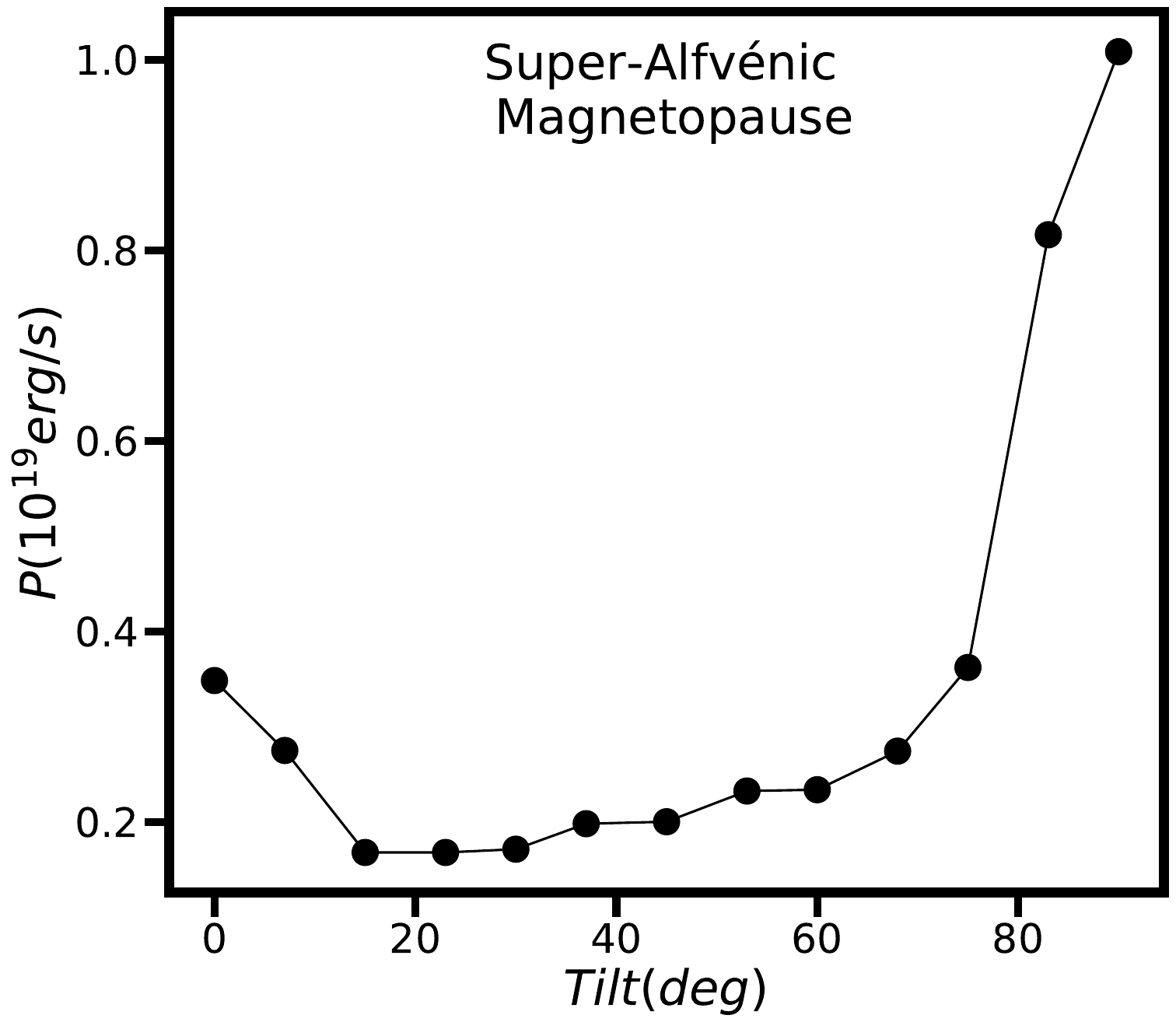}
    \includegraphics[width=.31\textwidth]{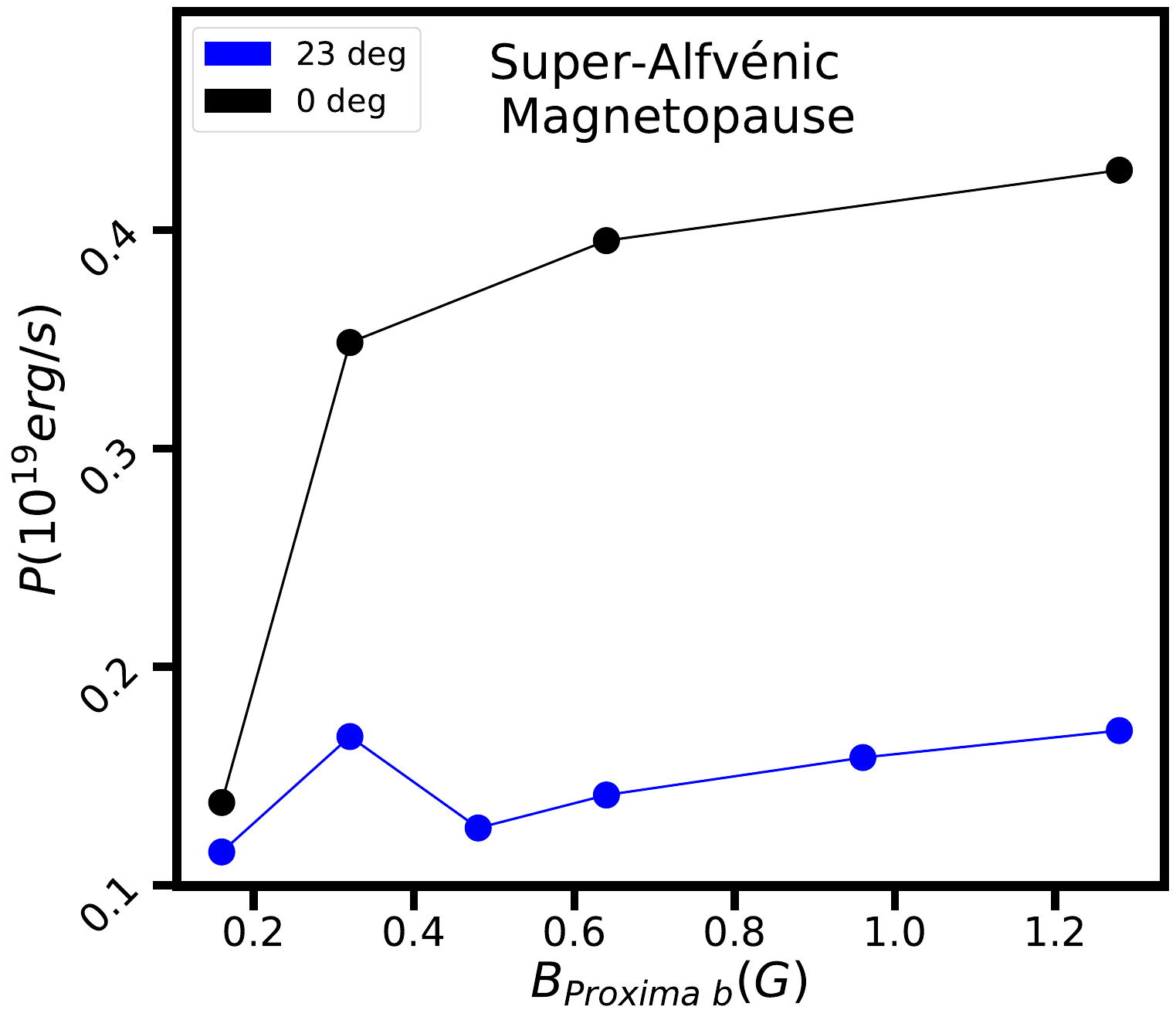}
  \end{minipage}
  \begin{minipage}{\textwidth}
    \centering
    \includegraphics[width=.31\textwidth]{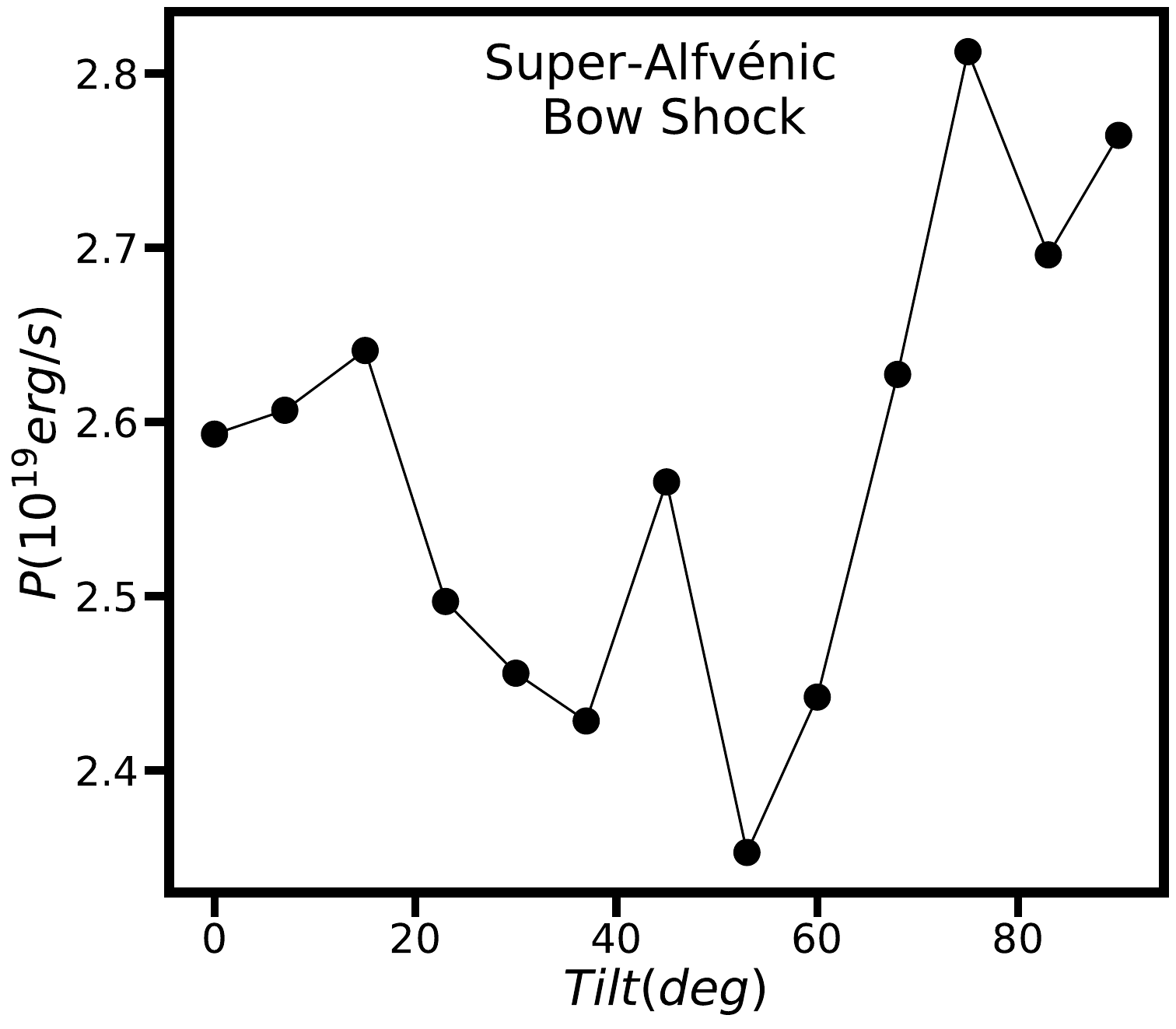}
    \includegraphics[width=.31\textwidth]{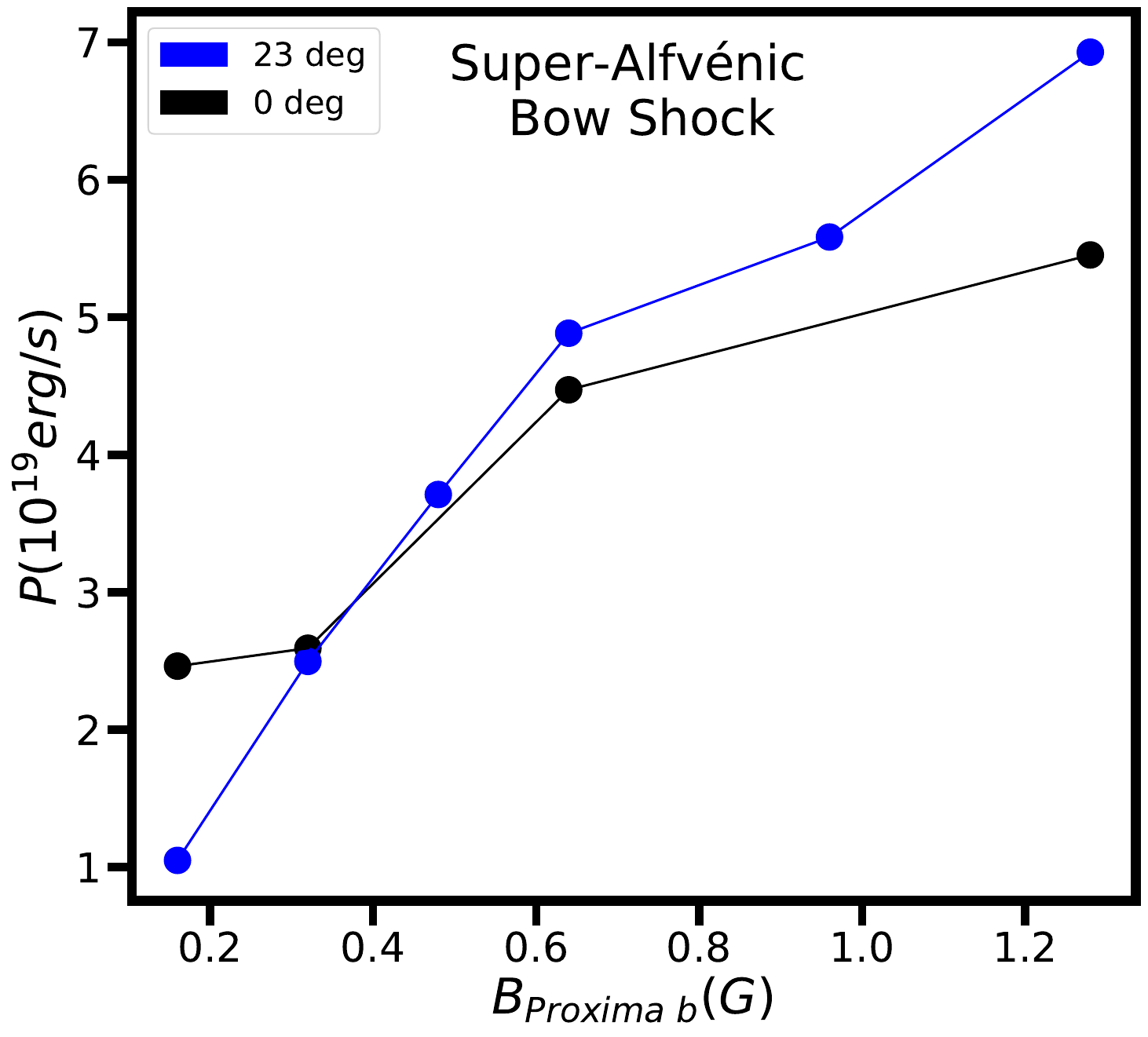}
  \end{minipage}
    \begin{minipage}{\textwidth}
    \centering
    \includegraphics[width=.31\textwidth]{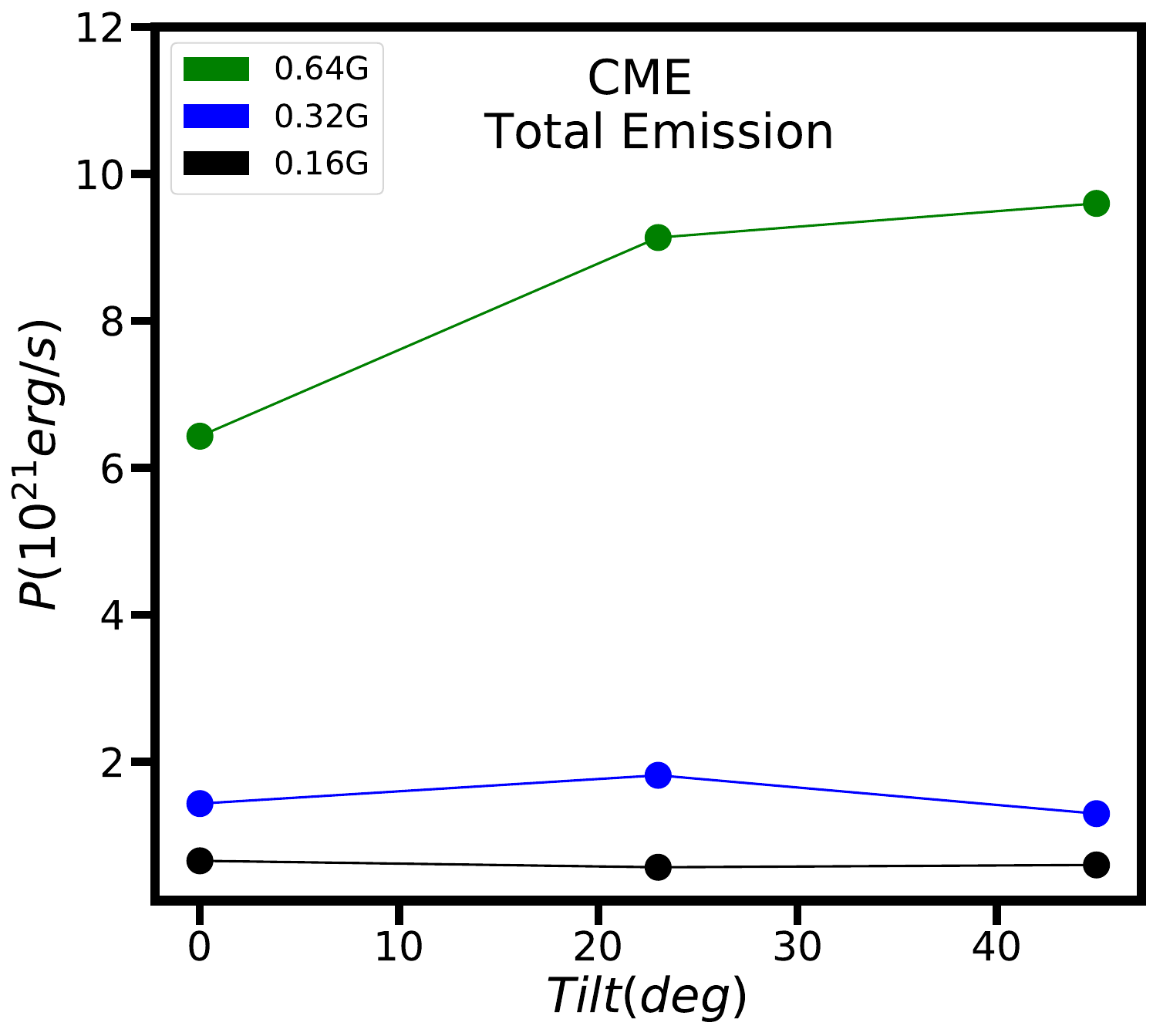}
    \includegraphics[width=.31\textwidth]{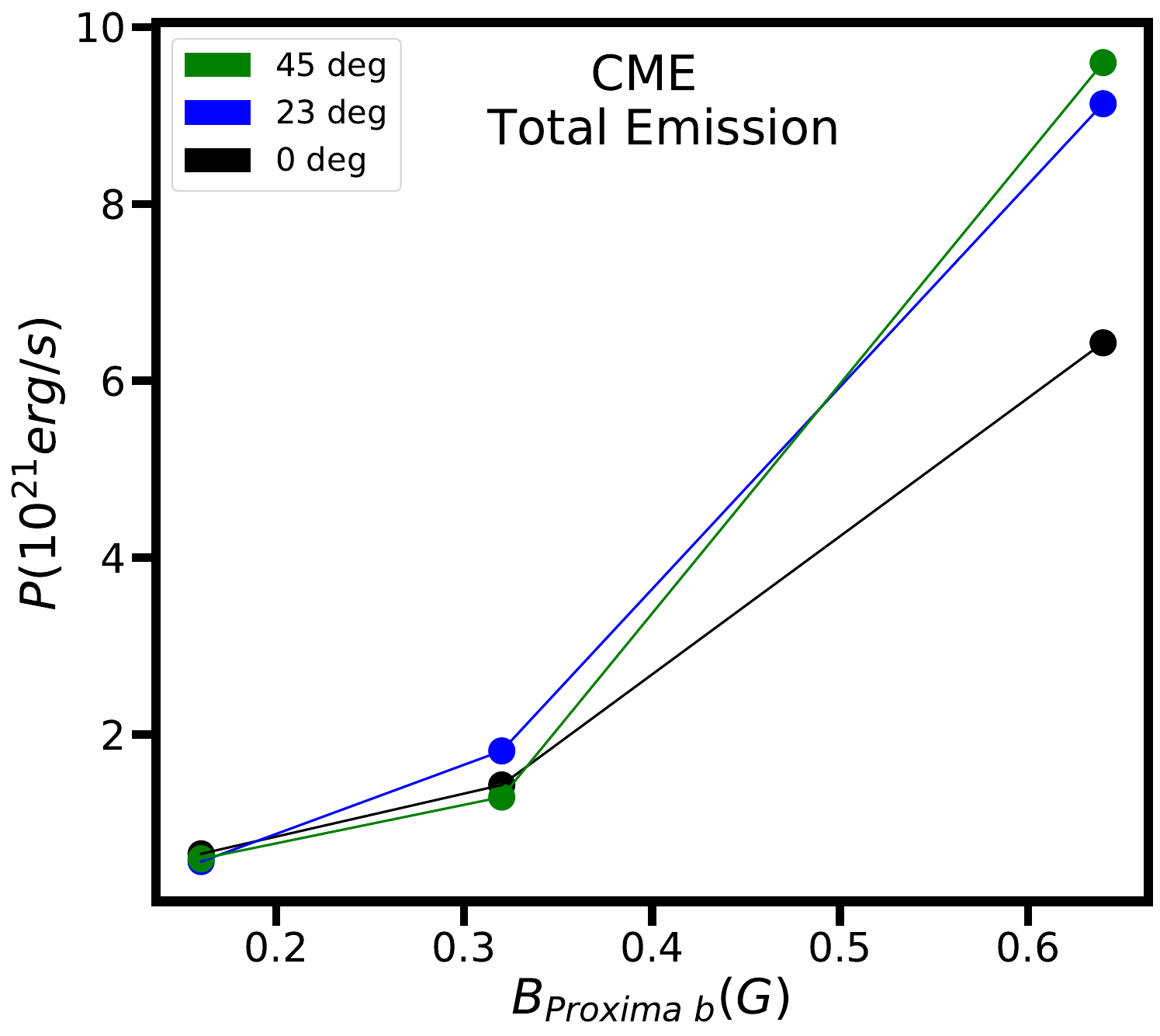}
  \end{minipage}
  \caption{
  Radio emission from the interaction of the stellar wind and the planetary magnetosphere 
of Proxima b for calm space weather conditions (sub-Alfvénic case--top panels; super-Alfvénic case--upper and lower middle panels) and extreme weather conditions (CME-like case--bottom panels), as a function of the tilt angle
and of the  magnetic field
of Proxima b. 
The top and upper middle panels correspond to the contribution from the magnetopause region, while the lower middle panels
correspond to the contribution from the bow shock in the super-Alfvénic scenario. For the CME-like case, the contribution is the total (magnetopause + bow shock).
}
\label{fig:radio-emission}
\end{figure*}

\begin{equation}
  \label{eq:radio-power}
 P_{R} = \beta P_{B} = \beta \int_{V} \vec{\nabla} \cdot
\frac{(\vec{\mathrm{v}}\wedge\vec{B})\wedge\vec{B}}{4\pi}dV, 
\end{equation} 

\noindent
where $P_{B}$ is the divergence of the magnetic Poynting flux associated with the
hot spots of energy transfer in the exoplanet day side, and $V$ is the volume
enclosed between the nose of the bow shock and the magnetopause.
We emphasise that we directly measure the radio emission, analysing our numerical results (see Appendix \ref{app:radio-emission-computation}), not by applying any approximate formula, other than the empirical factor $\beta$, as often done in the literature.

We note that the accelerated electrons that travel along the magnetic field lines generate cyclotron-maser emission near Proxima b (planetary emission)  or close to the star (we call this emission from star-planet interaction). The latter emission is expected to be produced only in the 
sub-Alfvénic regime, since one of the Alfvén wings is able to efficiently carry momentum and energy back to the star. In contrast,  planetary emission can be produced both in the sub- and super-Alfvénic regimes, since the radio-magnetic scaling law may hold in both cases. Our current analysis is focused on the emission from particles coming from the magnetopause reconnection regions, in particular the Poynting flux originated on the dayside, which likely dominates the overall radio emission budget, and therefore is enough to estimate the radio emission power, independently of the precise details of physical processes happening in the magnetosphere that lead to radio emission (e.g., Dungey cycle, electron acceleration.
Fig. \ref{fig:radio-emission} summarizes our results for both calm space weather and extreme space weather conditions. In that figure, we show the radio emitter power, $P_{\rm R}$, as a function of planet tilt (left panels) and of the magnetic field of Proxima b (right panels).

\subsection{Radio emission under calm space weather conditions}

The radio emission in the sub-Alfvénic scenario is produced only by charged particles coming from the magnetopause reconnection regions between the interplanetary magnetic field and the exoplanet magnetic field, as no bow shock is formed.  In the super-Alfvénic scenario, there is contribution from both the particles from the magnetopause and the bow shock,
which is linked to the interplanetary magnetic field pile-up and bending. 
We therefore obtained the individual contributions of particles coming from the magnetopause and the bow shock to facilitate comparisons, we summarize our results in Fig.~\ref{fig:radio-emission}.

The top panels of Fig. \ref{fig:radio-emission} correspond to the  analysis from the simulations for the sub-Alfvénic regime (see Table~\ref{tab:pluto-params} of Proxima b. Since no bow shock is formed, the dissipated power (and accelerated electrons) comes only from the magnetopause region. The top left panel shows the radio emission as a function of the planetary tilt.  The radio emission decreases slightly, from the case with no tilt to values of about $20^{\degree}$. For larger tilts, the radio emission monotonically increases, reaching approximately 2$\times 10^{19}$ \ergs, which are about two to three times larger than for medium tilt values. 

The decrease of the radio emission in the cases with low tilt values is due to a weakening of the reconnection between the IMF and the magnetic field of Proxima b  in the Southern region of the magnetopause.
As the tilt value increases, the reconnection region moves from the Southern 
to the equatorial region of Proxima b, which enhances the reconnection and
leads to larger radio emission. The value of the radio emission reaches a
maximum when the field lines of \Bimf\ and of \Bp\ are parallel in the nose of
the bow shock region, i.e., for a tilt of $90^{\degree}$.

\begin{figure}[htb]
  \centering
  \begin{minipage}{0.5\textwidth}
    \centering
    \includegraphics[trim={0cm 0cm 1cm 0},clip,width=.49\textwidth]{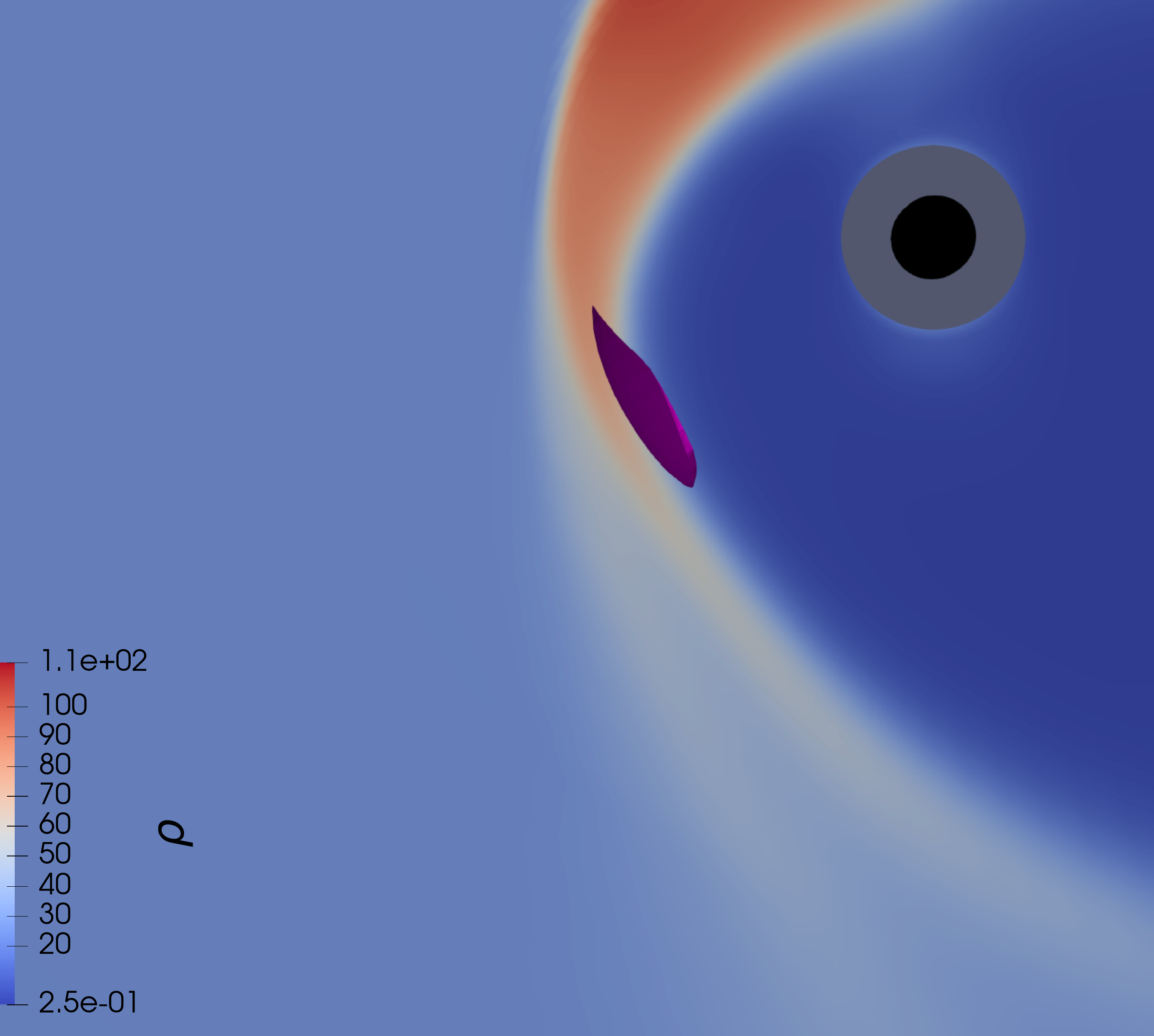}
    \includegraphics[trim={0cm 0cm 1cm 0},clip,width=.49\textwidth]{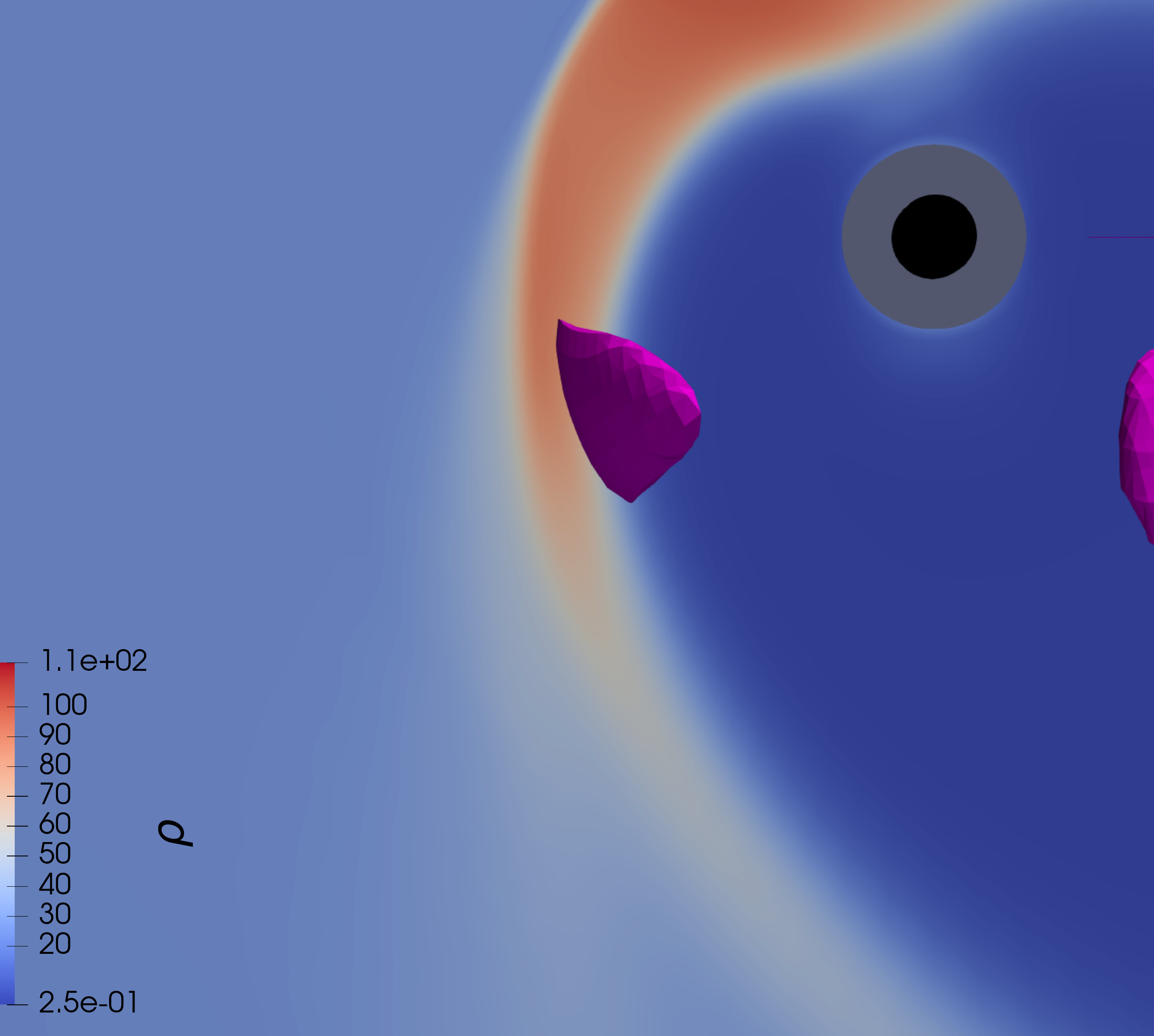}
  \end{minipage}
  \begin{minipage}{0.5\textwidth}
    \centering
    \includegraphics[trim={0cm 0cm 1cm 0},clip,width=.49\textwidth]{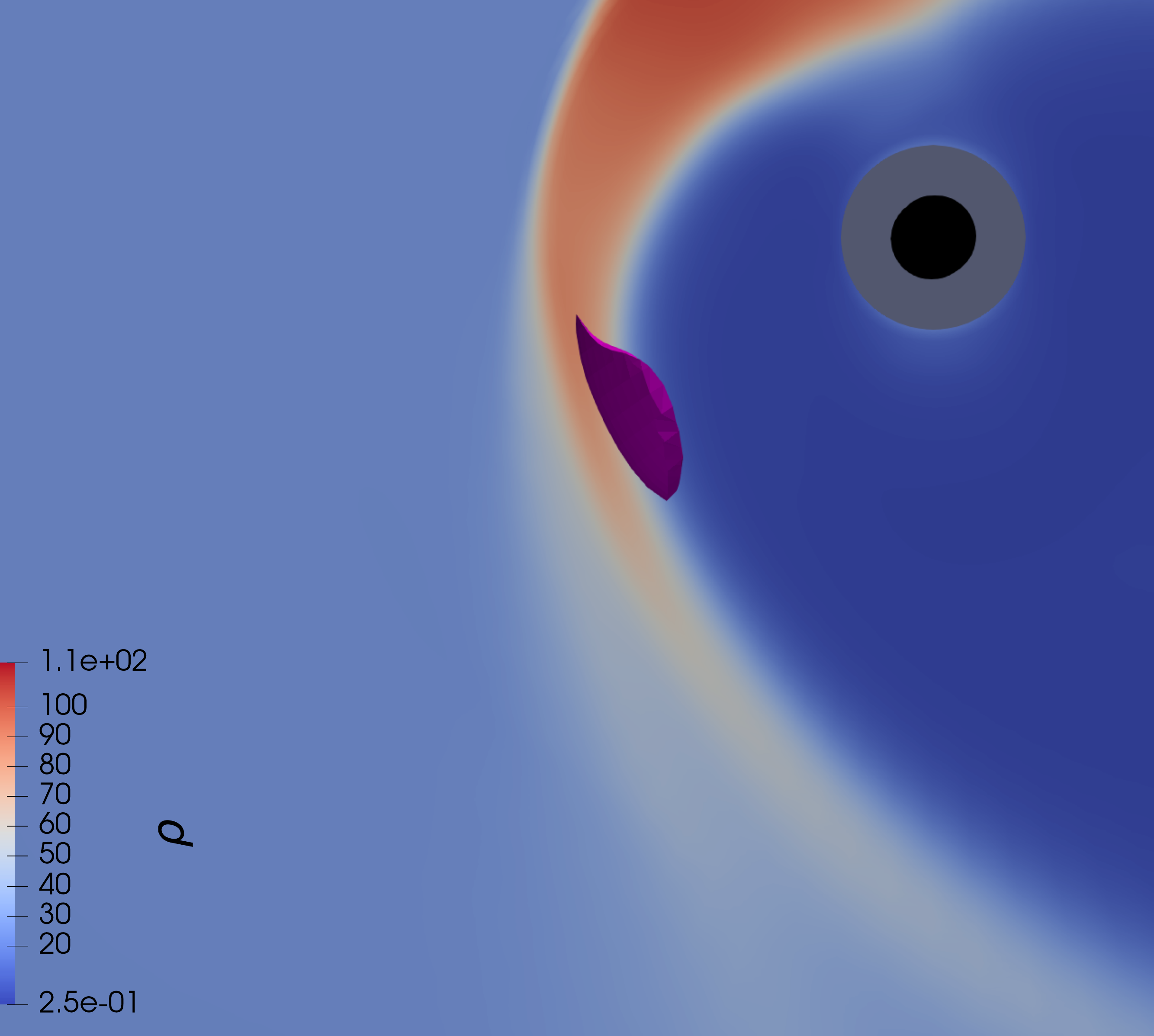}
    \includegraphics[trim={0cm 0cm 1cm 0},clip,width=.49\textwidth]{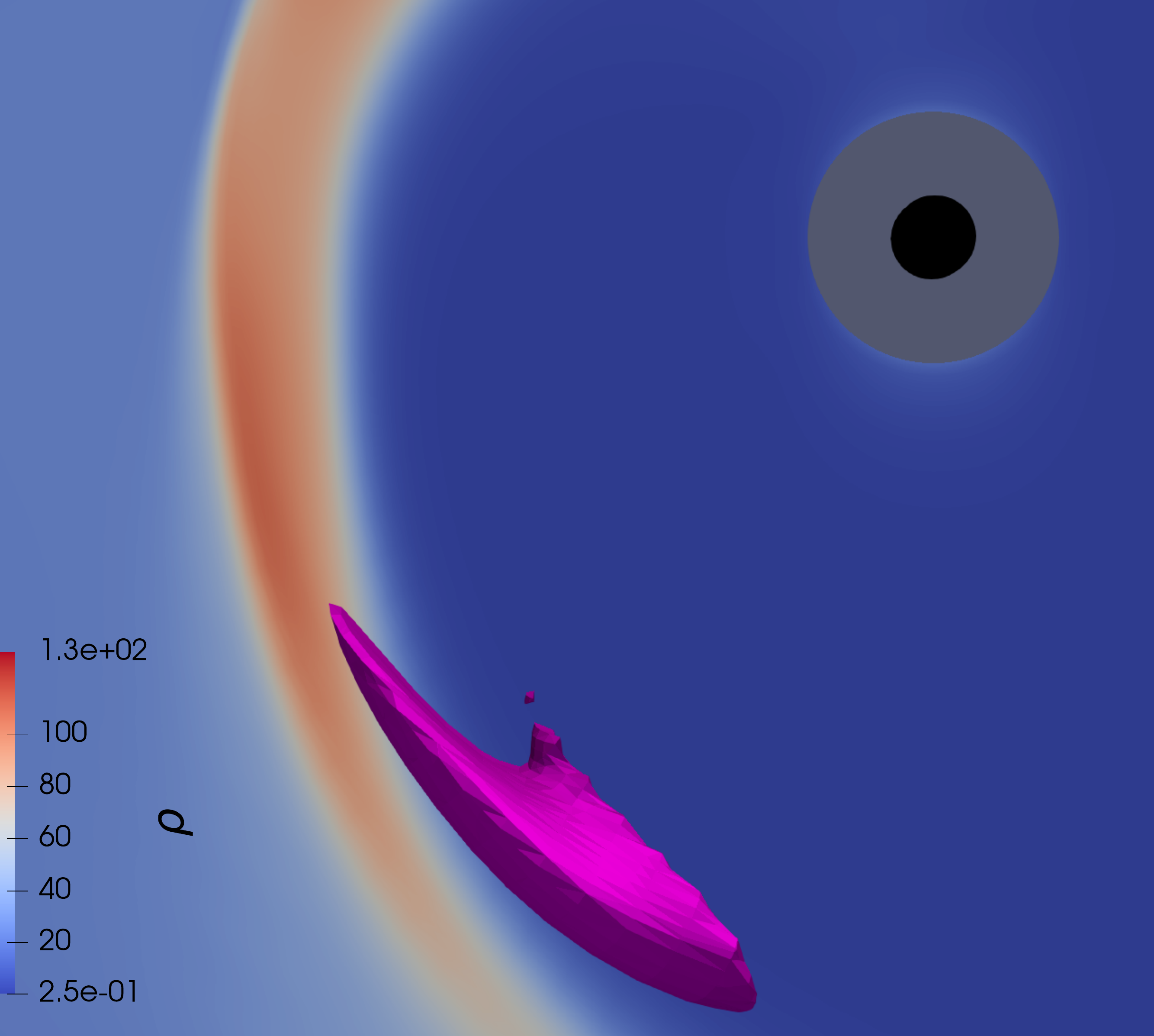}
  \end{minipage}
\caption{Close-up images of the region of interaction of the interplanetary magnetic
  field and that of Proxima b. The top panels highlight the effect of the
  tilt angle ($15^{\degree}$ and $37^{\degree}$ for the left and right panels,
  respectively, for a fixed value of \Bp\ = 0.32 G. The bottom panels show the
  effect of the planetary magnetic field on the reconnection region (0.32 G
  and 1.28 G for the left and right panels, for a fixed tilt of
  $23.5^{\degree}$). Note how the interaction region is clearly further away from the planet for the case of \Bp\ = 1.28 G.
  The reconnection region (pink colour) is defined as the region where the local magnetic field is $\leq 0.32$ mG, corresponding to seven times the normalized magnetic field value in our simulations. The planet Proxima b is shown as a black circle.  
} 
\label{fig:reconnection}
\end{figure}

The radio emission also increases steadily as a function of the exoplanetary
magnetic field (top right panel), as expected from Eq.~\ref{eq:radio-power}, reaching a maximum
value for $B_{\rm p} = 1.28$ G of $\sim 2.2 \times 10^{19}$ \ergs. The
increment of the radio emission with the magnetic field intensity of Proxima b
is linked to the generation of a wider reconnection region,  located further
away from the exoplanet surface, as shown in the bottom panels of Fig.~\ref{fig:reconnection}.
Note also how similar is the behaviour of the radio emission vs. exoplanetary
magnetic field for the cases with $i=0^{\degree}$ and $i=23.5^{\degree}$ (top panels of Fig.~\ref{fig:reconnection}). This is
due to the generation of wider reconnection regions as the magnetic field of
Proxima b increases, which reduces the impact of small variations of the tilt
on the radio emission as the magnetic field of Proxima is stronger. 

We show in the upper and lower middle panels of Fig. \ref{fig:radio-emission} the
expected radio emission in the super-Alfvénic regime, as a function of the
planetary tilt (left panels), where we set a value of \Bp\ = 0.32 G (Earth-like value) for all runs, and as a function of the
planetary magnetic field (right panels).  In this case, both a magnetopause and a bow
shock are formed, and those cases are shown in the upper and lower middle panels,
respectively. The overall trend of the radio emission from the magnetopause as
a function of the tilt angle (middle left panels) is similar to the
sub-Alfvénic case, although the emitted power is at most of $\simeq 10^{19}$
\ergs, about a factor of two smaller than that produced in the sub-Alfvénic case. The
radio emission in the super-Alfvénic cases is smaller because the IMF
intensity is half compared to the sub-Alfvénic case
(Table~\ref{tab:pluto-params}) and the pile-up of the IMF lines  is larger due
to the presence of the bow shock.

The radio emission of the magnetopause as a function of the Proxima b magnetic
field (upper middle right panel) increases with the magnetic field, by almost a
factor of three between the cases with \Bp\ = 0.16 and \Bp\ = 1.28 G, which is a
factor similar to that seen in the sub-Alfvénic cases.  Note also that the
radio emission is systematically smaller for a tilt equal to that of the
Earth, with respect to $i = 0^{\degree}$, in agreement with the result seen in
the upper middle left panel. The radio emission from the bow shock as a function of
the tilt angle does not show any evident trend.  It appears that the emitted
power is, within the uncertainties of the simulations, consistent with being
constant, at a level of about $\sim 2.6 \times 10^{19}$ \ergs, suggesting that
the planetary tilt plays a minor role, or no role at all, in the contribution
of the bow shock to the radio emission. Finally, the radio emission from the
bow shock as a function of \Bp\ (lower middle right panel) indicates that it monotonically increases with \Bp\,
reaching values of up to $7\times 10^{19}$ \ergs. The contribution from the
bow shock is about an order of magnitude larger than that of the magnetopause, and is
therefore the one that dominates the overall budget in the super-Alfvénic
regime. Further, there does not seem to be a strong difference in the radio
emission for the cases with $i= 0^{\degree}$ and $i = 23.5^{\degree}$, in
agreement with the study of the radio emission as a function of $i$.

There seems to be a trend in the dissipated radio power as a function of the planetary magnetic field (see panels for the sub-Alfvénic - magnetopause and super-Alfvénic - bow shock cases in Fig.~\ref{fig:radio-emission}). We fitted our data to a power law 
$P_{\rm R} = P_0 \times \left(B_{\rm p}/1 {\rm G}\right)^k$, where 
$P_0$ is the radio power for $B_{\rm p} = 1$ G.

For the sub-Alfvénic magnetopause emission, we obtain $P_0 = (2.1 \pm 0.1) \times 10^{19} $ \ergs\ and $k = 0.7 \pm 0.1$. For the super-Alfvénic bow shock emission, $P_0 = (5.7 \pm 0.6) \times 10^{19}$ \ergs\ and $k = 0.7 \pm 0.1$. For the super-Alfvénic magnetopause emission the fit does not show any clear correlation between $P_{\rm R}$ and $B_p$.

\subsection{Radio emission under CME-like conditions}

The bottom panels of Fig. \ref{fig:radio-emission} show the
predicted radio emission in a CME-like, super-Alfvénic scenario for Proxima b, as a function of the
planetary tilt (left) and as a function of the
planetary magnetic field (right).  Unlike in the super-Alfvénic case under calm space weather conditions discussed in the previous subsection, here we did not separate the contribution of the magnetopause from that of the bow
shock, and therefore give only the total contribution, which is anyway completely dominated by the bow shock. Note that the radio emission is more than two orders of magnitude larger than under calm weather conditions, whether in the sub-Alfvénic or super-Alfvénic regime. For a given magnetic field, there is no clear dependence with the tilt angle. On the other hand, the predicted radio emission as a function of the planetary magnetic field has an approximate $B_{\rm p}^2$ dependence, as could be expected from Eq.~\ref{eq:radio-power}.

The total radio power in the CME-like scenario shows also a  correlation with the planetary magnetic field. As in the calm space weather scenario, we fit the data to a power-law and find that $P_0 = (1.7 \pm 0.3)\times 10^{22}$ \ergs\ and  the index of the power-law is
$k = 1.9 \pm 0.2$. This value is significantly higher that in the case of calm space weather conditions, and compatible with $P_{\rm R} \propto B_{\rm p}^2$.

We note that in the CME-like scenario, both the bow shock and the magnetopause are closer to the planet than in the calm space weather scenario. In particular, the dynamic pressure of the stellar wind in the CME-like scenario is 30 times higher, which leads to a much more compressed configuration of the magnetosphere. The asymmetry induced by the IMF on the magnetic field of Proxima b is much smaller in this case, as the magnetic field close to Proxima b is more intense, and dominates over that of the IMF, despite it being 10 times larger than in the calm space weather scenario (super-Alfvénic). Therefore, those results might suggest 
that when the effect of the IMF on the magnetic topology of Proxima b is small in the CME-like scenario, the standard behaviour $P_{\rm R} \propto B_{\rm p}^2$ is recovered. If, on the other hand, the asymmetry induced by the IMF is large (calm space weather scenarios), then the dependence of 
$P_{\rm R}$ with $B_{\rm p}$ deviates from that standard behaviour.

In summary, the total radio emitted power obtained in the simulations is in the range of
$\sim (0.7-2)\times10^{19}$ \ergs\ and $\sim(1.2-7.5)\times10^{19}$ \ergs\ for
the sub-Alfvénic and super-Alfvénic regimes, respectively, under calm space weather conditions. Those values
correspond to radio flux densities in the range of $0.3 - 1.0$ mJy and up to about 3 mJy, for the sub-Alfvénic and super-Alfvénic regimes, respectively, assuming an isotropic emission ($\Omega = 4\pi$) and that the emission bandwidth is of 0.9 MHz, which is the value corresponding to the electron-cyclotron frequency of a planetary magnetic field of 0.32 G. 
If  Proxima b is subject to a more extreme space weather, e.g., a CME event, the predicted flux densities are more than about 100 times larger than when the planet is under calm average space weather conditions.

\citet{PerezTorres2021} measured radio emission values from the Proxima - Proxima b system from $\sim 200 \mu$Jy (steady emission) up to a few tens of mJy (bursty emission), using the Australia Telescope Compact Array at a frequency of $\sim 1.6$ GHz. While the coherent radio emission detected by \citet{PerezTorres2021} spread across a band of about $\sim$200 MHz and used a different efficiency factor in the conversion of the dissipated power to the radio power, those values broadly agree with the observations, which suggests that, if Proxima b is at times in a sub-Alfvénic regime, then radio emission from star-planet interaction can be detected. Not less important, our MHD simulations yield additional support to sub-Alfvénic interaction as a viable way for detecting signatures of exoplanets, albeit disentangling the emission from star-planet interaction from that of the star itself is challenging, and requests that the signal appears correlated with the orbital period of the (known) planet to confirm its star-planet interaction nature.
In the super-Alfvénic scenario the planetary emission can reach a few mJy and, if the planet is subject to a CME event, up to a few hundred mJy. Although those values are huge, the  emission frequency falls, even in its second harmonic, well below the cutoff frequency of the Earth ionosphere, and  therefore cannot be detected from Earth if the magnetic field of Proxima b is similar to the one of our planet. 

\section{Summary}
\label{sec:summary}

In this paper, we discussed the habitability of Proxima b, as well as the expected radio emission that arises from the interaction of the stellar wind magnetic field with that of the planet, by characterizing the magneto-plasma environment of Proxima b, when the planet is subject to either calm average space weather conditions, or to more extreme CME-like conditions.  We studied the role of the stellar wind and planetary magnetic field, and their mutual orientation, using the 3D MHD code PLUTO \citep{Mignone2007}, both for the sub-Alfvénic and super-Alfvénic regimes. We also predicted the radio emission generated  from the interaction between the
stellar wind of Proxima and the magnetosphere of its planet Proxima b, which
is relevant to guide radio observations aimed at unveiling planets. Compared to previous MHD simulations of the space weather around Proxima b, we measure the magnetopause standoff distance of Proxima b, \Rmp, directly from our simulations. This is unlike done in other works, where an approximate analytical expression is used, which can lead to values of \Rmp\ that largely depart from the true ones. Likewise, we  determine from our simulations the expected radio emission from the exoplanet dayside magnetosphere reconnection regions around Proxima b, which is relevant not only for the predictions regarding Proxima b, but also for essentially any Earth-like planet.

The main outcomes from this study are the following ones:

$-$ The magnetopause standoff distance, \Rmp, decreases with the tilt, $i$, and increases with \Bp\ (Fig.~\ref{fig:standoff}), both under calm and extreme space whether conditions. If Proxima b has a magnetic field similar to that of the Earth (\Bp=0.32 G) or larger, the shielding provided by the planetary magnetopause under calm space weather conditions is  more than enough for essentially any planetary tilt angle, $i$, but the most extreme ones, when $i$ is close to $90^{\degree}$. Even if the magnetic field is lower than that of the Earth, the magnetopause provides a good enough shielding, as long as the planetary tilt is similar to that of the Earth, or smaller. If Proxima b is subject to more extreme space weather conditions, e.g., a CME event from its host star, the planet is well shielded if the planet is an Earth-like planet (\Bp = 0.32 G, $i=23.5\degree$), or has a smaller tilt, or a larger planetary magnetic field.

$-$ 
We note that \Rmp\ is smaller in the 
sub-Alfvénic case, compared to the value for the super-Alfvénic case. This
paradoxical result appears because the dynamic pressure of the stellar wind is four times larger in the super-Alfvenic configurations.
On the other hand, the magnetic pressure of the IMF is four times larger in the sub-Alfvénic case, so the reconnection  intensity is stronger. This  
leads to an enhanced erosion of the magnetic field of Proxima b. 
Our simulations therefore imply that in the sub-Alfvénic case the effect of the IMF erosion on the magnetic field of Proxima is stronger than the effect of the enhanced magnetosphere compression in the
super-Alfvénic case.

$-$
We also determined the radio emission generated  from the interaction between the
stellar wind of Proxima and the magnetosphere of Proxima b,  which
is relevant to guide radio observations aimed at unveiling planets.
We find that, under calm space weather conditions,
the radio emission caused by the day-side reconnection regions can be as high
  as 7$\times10^{19}$ \ergs\ in the super-Alfvénic regime, and is on average
  almost an order of magnitude larger that the radio emission in the sub-Alfvénic
  case. This is due to the much larger contribution of the bow shock to the overall radio
  emission budget, which is not formed in the sub-Alfvénic regime. 
  We also find that the emission from the bow shock has a rather weak dependence on the tilt angle of the planet. If Proxima b is subject to extreme space weather conditions, the radio emission is more than two orders of magnitude larger than under calm space weather conditions. For a given planetary magnetic field, there is no clear trend of the expected radio emission with the tilt angle.
  On the other hand, we find that $P_{\rm R} \propto B_{\rm p}^{1.9 \pm 0.1}$. 
 This result might suggest the effect of the IMF on the magnetic topology of Proxima b is small in the CME-like scenario, leading to the standard scaling $P_{\rm R} \propto B_{\rm p}^2$, unlike in the calm space weather scenario, when the asymmetry induced by the IMF on the magnetic topology of the planet is large, which weakens the 
dependency of $P_{\rm R}$ on $B_{\rm p}$.

  The radio emission power from the interaction of the Proxima stellar wind with the magnetosphere of Proxima b should therefore result in
  flux densities at the distance of the Earth of up to a few mJy under calm space weather conditions (few hundred mJy, if the planet suffers the effects of CME-like events). However, the frequency of this radio emission falls below the ionosphere cutoff, and cannot therefore be detected from the Earth, but would require sensitive radio interferometric arrays beyond the ionosphere. On the other hand, this result yields expectations for a direct detection--from Earth--in radio of giant planets in close-in orbits, as those Jupiter-like planets, are expected to have magnetic fields large enough, so that their electron-cyclotron frequency falls above the ionosphere cutoff.

\begin{acknowledgements}
LPM and MPT acknowledge financial support through the Severo Ochoa grant CEX2021-001131-S and the National grant PID2020-117404GB-C21, funded by MCIU/AEI/ 10.13039/501100011033. JVR ackowleges support through project 2019-T1/AMB-13648, funded by the Comunidad de Madrid. LPM also acknowledges funding through grant PRE2020-095421, funded by MCIU/AEI/10.13039/501100011033 and by FSE Investing in your future.  We also acknowledge the Spanish Prototype of an SRC (SPSRC) service and support financed by the Spanish Ministry of Science, Innovation and Universities, by the Regional Government of Andalusia, by the European Regional Development Funds and by the European Union NextGenerationEU/PRTR.
PZ acknowledges funding from the ERC under the European Union’s Horizon 2020 research and innovation programme (grant agreement no. 101020459 - Exoradio).
\end{acknowledgements}

\bibliographystyle{aa}
\bibliography{References}

\begin{appendix}

\section{Upper ionosphere model}
\label{app:upper-ionosphere}

The upper ionospheric domain is located between $R = (2.2 - 2.5)\,R_{\rm E}$. The upper ionosphere model is based on \citet{Buchner2003}.
Below the lower boundary of the upper ionosphere, the magnetic field intensity is too high, thus the simulation time step is too small. In addition, a single-fluid MHD model cannot correctly reproduce magnetosphere regions such as the inner ionosphere or the plasma sphere because the kinetic effects are strong.

First, we calculate the field-aligned current ($J_{FAC}$) as
\begin{equation}
\label{eqn:1}
\vec{J_{FAC}} = \vec{J} - \vec{J_{\perp}}
,\end{equation}
where
\begin{equation}
\label{eqn:2}
\vec{J} = \frac{1}{4\pi}\vec{\nabla} \times \vec{B}
\end{equation}
\begin{equation}
\label{eqn:3}
\vec{J_{\perp}} = \vec{J} - \frac{J_{r} B_{r} + J_{\theta} B_{\theta} + J_{\phi} B_{\phi}}{|B|^{2}} \vec{B}
,\end{equation}
with $\vec{J}$ the plasma current, $\vec{J_{\perp}}$ the perpendicular component of the plasma current along the magnetic field line, $\mu_{0}$ the vacuum magnetic permeability, and $\vec{B}$ the magnetic field.
Next, we compute the electric field of the upper ionosphere model using the Pedersen conductance ($\sigma$) empirical formula,
\begin{equation}
\label{eqn:4}
\sigma = \frac{40 E_{0} \sqrt{F_{E}}}{16 + E_{0}^{2}}
,\end{equation}
with $E_{0} = k_{B} T_{e}$ the mean energy of the electrons, $F_{E} = n_{e} \sqrt{E_{0}/(2 \pi m_{e})}$ the energy flux, and $k_{B}$ the Boltzmann constant ($T_{e}$ and $m_{e}$ are the electron temperature and mass, respectively). Thus, the electric field ($E$) linked to the FAC is
\begin{equation}
\label{eqn:5}
\vec{E} = \sigma \vec{J_{FAC}}
.\end{equation}
Once we get the electric field, we compute the the velocity of the plasma in the upper ionosphere using the standard relation
\begin{equation}
\label{eqn:6}
\vec{v} = \frac{\vec{E} \times \vec{B}}{|B|^{2}}
.\end{equation}
We defined the plasma density in the upper ionosphere with respect to the Alfv\'{e}n velocity. More specifically, we fixed the module of the Alfv\'en velocity, $\mathrm{v}_{A}$, to control the simulation time step, so that the density profile, defined as
\begin{equation}
\label{eqn:7}
\rho = \frac{|B|^{2}}{4\pi\,v_{A}^{2}}
.\end{equation}
did not evolve during the simulation between $R = (2.2 - 2.5)\, R_{E}$. We note that, while we kept fixed the value of $\mathrm{v}_{A}$ in each simulation, depending on the specific simulation this value could be different, as a result of the different values of the IMF and of the stellar wind density. The values of $\mathrm{v}_{A}$  ranged from  $2.6 \cdot 10^{4}$ km/s to $5.0 \cdot 10^{4}$ km/s. 

We defined the plasma pressure in the upper ionosphere model with respect to the sound speed of the stellar wind, $c_{\rm sw}$, and at the inner boundary, $c_{\rm p}$),

\begin{equation}
\label{eqn:8}
p = \frac{n}{\gamma} \left[ \frac{(c_{\rm p} - c_{\rm sw})(r^{3}-R_{s}^{3})}{R_{un}^{3}-R_{s}^{3}} + c_{\rm sw} \right]^{2}
,\end{equation}
with $\gamma = 5/3$ the polytropic index, $c_{p} = \sqrt{\gamma K_{B} T_{p}/m_{p}}$ with $T_{p}$ the plasma temperature at the inner boundary, and $c_{sw} = \sqrt{\gamma K_{B} T_{sw}/m_{p}}$ with $T_{sw}$ the stellar wind temperature.

We defined the initial model conditions of the plasma density and pressure so as to have a smooth transition between the upper ionosphere and the simulation domains. During the simulation, the pressure and density gradients increase because we kept fixed the density and pressure profiles inside the inner ionosphere, but evolved freely in the simulation domain. The reaction of the system during the early stages of the simulation is to feed plasma towards the simulation domain to compensate for the increase of the pressure and density gradients. This generates an outward plasma flux that saturates when the inner magnetosphere reaches a steady state. Henceforth, the plasma flows are driven by the balance between the stellar wind injection inside the inner magnetosphere, and the plasma streams towards the planet surface.

We note that the present model has been bench-marked against several codes, following the analysis performed by \citet{Samsonov2016}, which was dedicated to the global structures of the Earth magnetosphere for quiet space weather conditions. Namely, we performed a simulation using the same parameters as in the original benchmark study: $n = 5$ cm$^{-3}$, $V_{x} = -400$ km/s, $T = 2 \cdot 10^{5}$ K, 
$B_{y} = -B_{x} = 35\, \mu$G , and $B_{z} = 0\, \mu$G . 
The location of the magnetopause is $R_{x}/R_{E} = 10.7$, $R_{y}/R_{E} = 16.8$, $R_{-y}/R_{E} = 16.6,$ and $R_{z}/R_{E} = 14.9$. The model prediction and the benchmark study agree reasonably well. In addition, the electric field in the simulation domain is also consistent with the simulations in \citet{Samsonov2016} 
near the bow shock. The module of the electric field predicted inside the magnetosphere is similar to Cluster spacecraft observations during the magnetopause crossing on $30/02/2002$ \citep{Keyser2005}. The electric field measured in the current sheet and magnetosheath is an order of magnitude higher than the simulations because the IMF module is $10$ times larger during the Cluster magnetopause crossing. When the simulation was performed using a southward IMF with 
$|B|=500\,\mu$G and $P_{d} = 5$ nPa, similar to the space 
weather conditions during Cluster magnetopause crossing, the predicted electric field is $15-30$ mV/m in the current sheet and magnetosheath region. This is similar to Cluster spacecraft observations.

We also performed another two simulations using the same SW parameters, but for northward and southward IMF orientations with $|B_{z}|=3$ nT, identifying the displacement of the magnetopause location defined as $\Delta R/R_{E} = {\rm northward}(R)/R_{E} - {\rm southward}(R)/R_{E}$: $\Delta R_{x}/R_{E} = 0.2$, $\Delta R_{y}/R_{E} = 0.1$ and $\Delta R_{z}/R_{E} = -1.0$. As in the previous case, we also found reasonble agreement with the benchmark case.

Next, we compared that model with the Carrington-like event analysed by \citet{Ridley2006}, who  identified a magnetopause standoff distance of $R/R_{E} = 2$ (equal to the lower boundary of the simulation domain) for the parameters $n = 750$ cm$^{-3}$, $V_{x} = -1600$ km/s ($P_{d}=1600$ nPa), $T = 3.5 \cdot 10^{7}$ K, $B_{x} = 1.5$ mG, $B_{y} = 1.7$ mG, and $B_{z} = 2.0$ mG. This model cannot be used to simulate space weather conditions leading to a magnetopause standoff distance smaller $R/R_{E} = 2.5$, although the extrapolation of the model results predicts $R/R_{E} \approx 1.22$ if $P_{d}=1600$ nPa and $B_{z} = 2.0$ mG (pure southward IMF orientation).

Finally, we note that the electric field in the upper ionosphere domain remains almost unchanged during the simulation because we had fixed the density profile. The radial electric field inside the upper ionosphere (Northern hemisphere at $R/R_{\rm E} = 3.1$) 
shows a reasonable agreement with respect to other models and satellite measurements \citep{Shume2009,Watanabe2014}.

FAC intensity and orientation values are in the range of the observations and modeling data (from $nA / m^2$ to several $\mu A m^2$ regarding space weather conditions) 
\citep{Weimer2001,Waters2001,Ritter2013,Bunescu2019,Zhang2020}

\section{Calculation of the radio emission from the interaction regions}
\label{app:radio-emission-computation}

We calculated the power dissipated in the interaction regions between the stellar wind and the magnetosphere at the exoplanet dayside, using Eq.~\ref{eq:radio-power}. 
In the (calm weather) sub-Alfvénic scenario, no bow shock is formed, so the only region contributing to the radio emission is the magnetopause. In the super-Alfvénic case (both under calm space weather conditions, or under more extreme, CME-like conditions), a bow shock is also formed, so there are two regions contributing to the dissipation of the deposited energy.

To perform those calculations, we defined the integration regions as accurately as
possible. For example, the magnetopause emission takes place around the
reconnection region. We excluded the magneto-tail zone from it, as it was not
included in the numerical modelling. Also, we put special care, so that the integration region did not fall
too close to the planet, as this could cause artificial emission near the boundaries of the simulation.  In the cases when a bow shock is formed, the
integration region coves all the space with high density, arising from
the collision of the stellar wind with the magnetic field of the planet.  We ensured that the
inner boundary was similar to the outer boundary of the magnetopause, so as to
avoid overlapping emission. 

\begin{figure}[h]
\centering
\resizebox{\hsize}{!}{\includegraphics[width=\columnwidth]{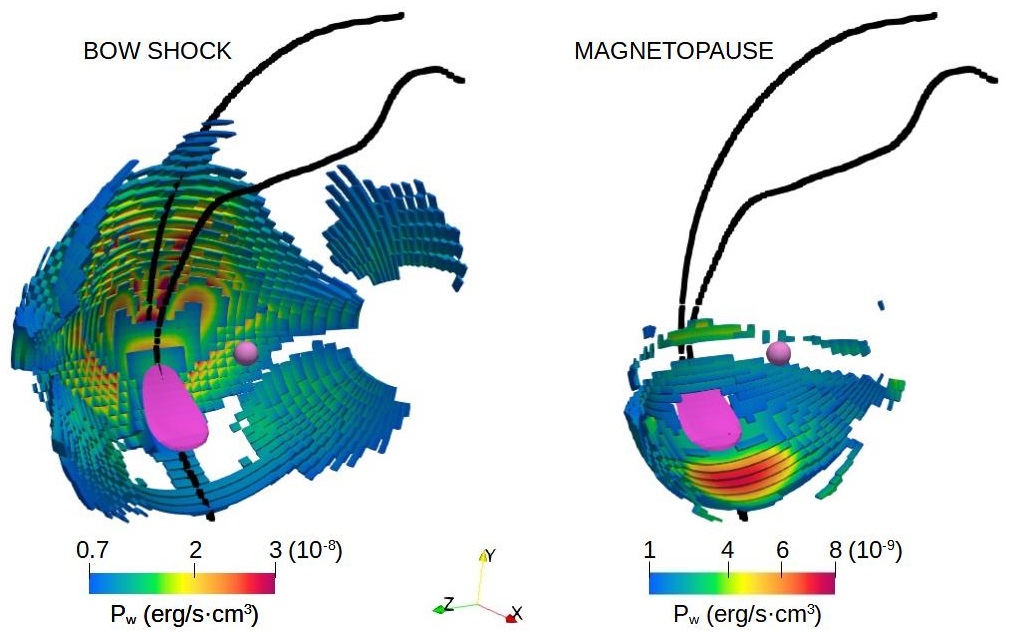}}
\caption{
Contribution to the radio emission from the bow shock (left) and magnetopause
(right) regions for one of our simulations. The black solid lines indicate the bow shock position, and the area coloured in purple represents the magnetic reconnection region.
} 
\label{fig:bowshock-magnetopause}
\end{figure}

We illustrate in Fig.~\ref{fig:bowshock-magnetopause} the different contributions from the
bow shock and the magnetopause for a super-Alfvénic case, after having carried out the steps outlined.

\end{appendix}

\end{document}